\documentclass[%twocolumn,showpacs,
preprintnumbers]{revtex4}
\usepackage{graphicx}
\begin{document}
\draft
\preprint{\vtop{{\hbox{YITP-05-75}\vskip-0pt
                 \hbox{KANAZAWA-05-17}\vskip-0pt
%                 \hbox{hep-ph/05?????}
}}}
\thispagestyle{empty}
\title{On charm scalar resonances\footnote{Invited talk at 
the workshop on {\it Resonances in QCD}, July 11 -- 15, 2005, ECT*, 
Trento, Italy}
}
\author{K. Terasaki}%\email{terasaki@yukawa.kyoto-u.ac.jp; 
%terasaki@hep.s.kanazawa-u.ac.jp}
\affiliation{\hspace{-3mm}
Yukawa Institute for Theoretical Physics, 
Kyoto University, Kyoto 606-8502, Japan, \\
Institute for Theoretical Physics, Kanazawa University, 
Kanazawa 920-1192, Japan
}
%\date{October ??, 2005}
\thispagestyle{empty}
%%%%%%%%%%%%%%%%%%%%%%%%%%%%%%%%%%%%%%%%%%%%%%%%%%%%%%%%%%%%%%%%%%%%%%
\begin{abstract}
A brief overview of charm scalar resonances is given. It is demonstrated 
that experimental data on the $D_s^{*+}\gamma$ and $D_s^+\pi^0$ 
decays of $D_{s0}^+(2317)$ favor its assignment to the $I_3=0$ component 
of the iso-triplet $[cn][\bar s\bar n]_{I=1}$ mesons. 
The observed broad bump just below a large peak arising from the 
well-known tensor $D_2^*$ meson in a $D\pi$ channel is also studied.  
\end{abstract}
%%%%%%%%%%%%%%%%%%%%%%%%%%%%%%%%%%%%%%%%%%%%%%%%%%%%%%%%%%%%%%%%%%%%%%%%%%%%

\vskip 0.5cm
%\pacs{%PACS number(s)
%14.40.Lb, 13.25.Ft}
\maketitle
%%%%%%%%%%%%%%%%%%%%%%%%%%%%%%%%%%%%%%%%%%%%%%%%%%%%%%%%%%%%%%%%%%%%%%%%%%%%
\section{Introduction}
%%%%%%%%%%%%%%%%%%%%%%%%%%%%%%%
A charm-strange scalar meson $D_{s0}^+(2317)$ was discovered by the 
BABAR collaboration~\cite{BABAR-D_s}, and its existence has been confirmed 
by the BELLE~\cite{BELLE-D_s}, CLEO~\cite{CLEO-D_s} and 
FOCUS~\cite{FOCUS-D_s} collaborations. Its mass and width are now compiled 
as 
%%%%%%%%%%%%%%%%%%%%%%%%%%%%%%%%%%%%%%%%%%%%%%%%%%%%%%%%%%%%%%%%%%%%%%%%%%%%  
\begin{equation}
m_{D_{s0}}=2317.4 \pm 0.9\,\,{\rm MeV}, \quad 
\Gamma_{D_{s0}} < 4.7\,\,{\rm MeV}   
        \label{eq:D_s-data}
\end{equation}
%%%%%%%%%%%%%%%%%%%%%%%%%%%%%%%%%%%%%%%%%%%%%%%%%%%%%%%%%%%%%%%%%%%%%%%%%%%%
by the particle data group 2004 (PDG04)~\cite{PDG04}. It has been observed 
in the $D_s^+\pi^0$ channel but no signal in the radiative
$D_s^{*+}\gamma$ 
has been detected at the BABAR and at the CLEO, so that the CLEO 
collaboration~\cite{CLEO-D_s} has given a severe constraint, 
%%%%%%%%%%%%%%%%%%%%%%%%%%%%%%%%%%%%%%%%%%%%%%%%%%%%%%%%%%%%%%%%%%%%%%%%%%%
\begin{eqnarray}
{\Gamma(D_{s0}^+(2317)\rightarrow D_{s}^{*+}\gamma)
\over 
\Gamma(D_{s0}^+(2317)\rightarrow D_{s}^{+}\pi^0)}\Biggl|_{\rm CLEO} 
< 0.059. 
                                                   \label{eq:CLEO}
\end{eqnarray}
%%%%%%%%%%%%%%%%%%%%%%%%%%%%%%%%%%%%%%%%%%%%%%%%%%%%%%%%%%%%%%%%%%%%%%%%%
%The BELLE collaboration~\cite{BELLE-radiative} provided another constraint 
%on the same ratio, 
%%%%%%%%%%%%%%%%%%%%%%%%%%%%%%%%%%%%%%%%%%%%%%%%%%%%%%%%%%%%%%%%%%%%%%%%%%
%\begin{eqnarray}
%R(D_{s0}^+(2317)) \equiv 
%{\Gamma(D_{s0}^+(2317)\rightarrow D_{s}^{*+}\gamma)
%\over 
%\Gamma(D_{s0}^+(2317)\rightarrow D_{s}^{+}\pi^0)}\Biggl|_{\rm BELLE} 
%< 0.18
%                                                  \label{eq:BELLE-ratio}
%\end{eqnarray}
%%%%%%%%%%%%%%%%%%%%%%%%%%%%%%%%%%%%%%%%%%%%%%%%%%%%%%%%%%%%%%%%%%%%%%%%%
%from analysis of $D_{s0}^+(2317)$ productions through $B$ decays. 

For non-strange charm scalar mesons $D_0$, two independent observations of 
a broad bump just below a large peak arising from the well-known tensor 
$D_2^*$ in each of $(D\pi)^{0,+}$ mass distributions have been 
reported~\cite{BELLE-D_0,FOCUS-D_0} and it has been interpreted as the 
conventional scalar $D_0^{*}\{c\bar n\},\,(n=u,d)$ meson. 
Their measured masses and widths are 
%%%%%%%%%%%%%%%%%%%%%%%%%%%%%%%%%%%%%%%%%%%%%%%%%%%%%%%%%%%%%%%%%%%%%%%%%%
\begin{equation}
m_{D_{0}^+}=2308 \pm 60\,\,{\rm MeV}, \quad 
\Gamma_{D_{0}^+} =276 \pm 99\,\,{\rm MeV}
\end{equation}
%%%%%%%%%%%%%%%%%%%%%%%%%%%%%%%%%%%%%%%%%%%%%%%%%%%%%%%%%%%%%%%%%%%%%%%%%%
by the BELLE collaboration and  
%%%%%%%%%%%%%%%%%%%%%%%%%%%%%%%%%%%%%%%%%%%%%%%%%%%%%%%%%%%%%%%%%%%%%%%%%%
\begin{eqnarray}
&& m_{D_{0}^+}=2407 \pm 56\,\,{\rm MeV}, \quad 
\Gamma_{D_{0}^+} = 240\pm 114\,\,{\rm MeV},    \nonumber\\
&&m_{D_{0}^0}=2403 \pm 49\,\,{\rm MeV}, \quad 
\Gamma_{D_{0}^0} =283\pm 58\,\,{\rm MeV}
\nonumber
\end{eqnarray}
%%%%%%%%%%%%%%%%%%%%%%%%%%%%%%%%%%%%%%%%%%%%%%%%%%%%%%%%%%%%%%%%%%%%%%%%%%
by the FOCUS. In spite of the large difference between the central values 
of $m_{D_0}$ by the BELLE and by the FOCUS, we consider that they are of 
the same origin because they are consistent with each other within their 
large errors. We will discuss these broad bumps later. 

The mass of $D_{s0}^+(2317)$ was considerably lower than theoretical 
predictions of mass of the conventional scalar $\{c\bar s\}$ by the 
potential model~\cite{GK,potential}, the lattice 
QCD~\cite{quench,Bali,UKQCD}, %the QCD sum rule~\cite{QCDSR}, 
etc. Therefore, many authors have tried to assign it to various hadron 
states as listed in Table 1, where 
(a) the conventional scalar $\{c\bar s\}$~\cite{DGG} which is 
(b) the chiral partner of $D_s^+$~\cite{chiral,BH} in the heavy charm 
quark picture, prior to the observation of $D_{s0}^+(2317)$, and 
(c) a mixed state of the scalar $\{c\bar s\}$ and a $DK$ 
molecule~\cite{BCL}, 
(d) a mixed state of the scalar $\{c\bar s\}$ and an iso-singlet four-quark 
meson~\cite{BPP}, 
(e) an iso-singlet 
four-quark state~\cite{CH}, 
(f) an iso-triplet four-quark meson~\cite{Terasaki-D_s}, and 
(g) a dynamically generated resonance~\cite{BR,Lutz,atom}, etc., after 
the observation. 
The ratios, $R(D_{s0}^+)$ and $R(D_{s}^{*+})$, in the table are given by 
%%%%%%%%%%%%%%%%%%%%%%%%%%%%%%%%%%%%%%%%%%%%%%%%%%%%%%%%%%%%%%%%%%%%%%%%%%
\begin{equation}
R(D_{s0}^+)={\Gamma(D_{s0}^+\rightarrow D_s^{*+}\gamma) 
\over \Gamma(D_{s0}^+\rightarrow D_s^{+}\pi^0)}  \quad{\rm and}\quad          
R(D_s^{*+})={\Gamma(D_{s}^{*+}\rightarrow D_s^{+}\gamma) 
\over \Gamma(D_{s}^{*+}\rightarrow D_s^{+}\pi^0)}.            
\end{equation}
%%%%%%%%%%%%%%%%%%%%%%%%%%%%%%%%%%%%%%%%%%%%%%%%%%%%%%%%%%%%%%%%%%%%%%% 
For their experimental data, we take 
%%%%%%%%%%%%%%%%%%%%%%%%%%%%%%%%%%%%%%%%%%%%%%%%%%%%%%%%%%%%%%%%%%%%%%%
\begin{equation}
R(D_{s0}^+)_{\rm exp}< 0.059 \quad{\rm and}\quad 
%%%%%%%%%%%%%%%%%%%%%%%%%%%%%%%%%%%%%%%%%%%%%%%%%%%%%%%%%%%%%%%%%%%%%%%%%%
%\begin{equation}
R(D_s^{*+})^{-1}_{\rm exp}    %=0.062\pm 0.026.   
=0.062\pm 0.006\pm 0.005
                                         \label{eq:constraints}
\end{equation}
%%%%%%%%%%%%%%%%%%%%%%%%%%%%%%%%%%%%%%%%%%%%%%%%%%%%%%%%%%%%%%%%%%%%%%%%%% 
The former has been provided by the CLEO collaboration~\cite{CLEO-D_s} 
as mentioned before and the latter has been improved recently by the 
BABAR~\cite{BABAR-Radiative}. 
Models which cannot satisfy both of these data should be excluded. 

The ratios $R(D_{s0}^+)$ for $D_{s0}^+$ as the chiral partner of $D_s^+$ 
and as a $DK$ molecule have been calculated in Ref.~\cite{MS} as listed
in Table~1. The assignment to the $DK$ molecule may be excluded because  
the result of $R(D_{s0}^+)\sim 3$ which is far beyond the experimental 
upper bound in Eq.~(\ref{eq:CLEO}). In this type of models, one would have 
an additional broad resonance (above the $DK$ threshold) whose dominant 
component is the scalar $\{c\bar s\}$. As seen in Table~1, the assignment 
to the conventional scalar or the chiral partner of $D_s^+$ seems to be 
difficult to satisfy $R(D_{s0}^+)_{\rm exp}$ and $R(D_s^{*+})_{\rm exp}$ 
simultaneously. When $D_{s0}^+(2317)$ is assigned to the iso-singlet 
$\tilde D_{0s}\sim \{cn\bar n\bar s\}_{I=0}$ as in the case (e), 
it should be observed as a narrow peak on a broad ($\gtrsim 100$ MeV) 
bump arising from $\tilde D_{1s}^+$ in the $D_s^+\pi^0$ mass distribution 
if $\tilde D_{1s}^+$ is sufficiently produced. Besides, the conventional 
scalar $\{c\bar s\}$ might be observed as an additional (broad) resonance 
in the $DK$ channel, if its mass is higher than the $DK$ threshold. 
In the case (g), $D_{s0}^+(2317)$ has been considered as one 
of dynamically generated resonances. However, in this type of theories, 
their radiative decays have not yet been investigated. 
%%%%%%%%%%%%%%%%%%%%%%%%%%%%%%%%%%%%%%%%%%%%%%%%%%%%%%%%%%%%%%%%%%%%%%%%%%%%
%\newpage
%%%%%%%%%%%%%%%%%%%%%%%
\begin{center}
\begin{quote}
{Table~1. 
Various assignments of $D_{s0}^+(2317)$ and tentative comments on them. 
$R(D_{s0}^+)$ and $R(D_{s}^{*+})$ are given in the text. 
}
\end{quote}
\vspace{3mm}

\begin{tabular}
{| l | l |}
\hline
Assignments \qquad&\quad Comments\\   
\hline
\begin{tabular}{l}
Scalar {$\{c\bar s\}$}\hspace{45mm}(a)   \\
$\chi$-partner of  {$D_s^+$} \hspace{36mm}(b) 
\end{tabular}
& 
$\left\{\begin{tabular}{l}
{$\displaystyle{R(D_{s0}^+)\quad\,\,\simeq 0.13}$} \hspace{33mm} 
($\ddagger$)  \\
\hspace{17mm}$\simeq 0.08$ \hspace{33mm} ($\dagger$)\\
{$\displaystyle
{R(D_{s}^{*+})^{-1}\simeq 0.018}$} \hspace{31mm} ($\dagger$)\end{tabular}
\right.$
\\
\hline
\begin{tabular}{l}
Mixed state of  {$\{c\bar s\}$} 
and  {$DK$} molecule, \hspace{1.5mm}(c)  
\\
 \hspace{33mm} or {$\{cq\bar q\bar s\}_{I=0}$}  \hspace{6.5mm}  (d) 
\end{tabular}
& 
\begin{tabular}{l}
\hspace{2mm}
{$\displaystyle{R(D_{s0}^+)\,\,\,\,  \sim \,3}$} 
($\ddagger$) and Broad $D_{s0}^{*+}$ 
\\
\\
\end{tabular}
\\
\hline 
  {
$\tilde D_{0s}^+\sim\{cn\bar n\bar s\}_{I=0}$} \hspace{34.5mm}(e)
& \quad\begin{tabular}{l}
Narrow  {$\tilde D_{0s}^+$}, 
broad  {$\tilde D_{1s}^+$} (and $D_{s0}^{*+}$) 
\end{tabular}\\
\hline
  {
$\hat F_I^+\sim [cn][\bar s\bar n]_{I=1}$} \hspace{34mm} (f) 
& \quad\begin{tabular}{l} 
Narrow {$\hat F_I,\,\hat D,\,\hat D^s;
\,\hat F_0^+;\,\hat E^0$} and broad $D_{s0}^{*+}$\\
{$\displaystyle{R(D_{s}^{*+})^{-1} \sim 0.06}$}, 
{$\displaystyle{R(D_{s0}^+) \sim 0.005}$}
\end{tabular}\vspace{0mm}\\
\hline
 Dynamically generated \hspace{27mm}(g)
&
\quad\,{$R(D_{s0}^+)= ?$}, \quad {$R(D_s^{*+})^{-1}= ?$}
\\
\hline
\end{tabular}    
\end{center}
%%%%%%%%%%%%%%%%%%%%%%%%%%%%%%%%%%%%%%%%%%%%%%%%%%%%%%%%%%%%%%%%%%%%%%%%%%
\hspace{35mm}(a) in \cite{DGG}, (b) in \cite{chiral,BH}, 
(c) in \cite{BCL}, (d) in \cite{BPP}, (e) in \cite{CH}, 
(f) in \cite{Terasaki-D_s}, \vspace{-3mm}\\

\hspace{30mm}  (g) in \cite{atom,BR,Lutz}; 
($\ddagger$) in \cite{MS}, ($\dagger$) in \cite{BEH}
\vspace{5mm}\\
%%%%%%%%%%%%%%%%%%%%%%%%%%%%%%%%%%%%%%%%%%%%%%%%%%%%%%%%%%%%%%%%%%%%%%%%%%

In this article, we study charmed scalar four-quark mesons and their 
two-body decays, and then the decays of charm-strange scalar mesons into 
$D_s^+\pi^0$ and $D_s^{*+}\gamma$ final states in consistency with the  
$D_s^{*+}\rightarrow D_s^+\pi^0$ and $D_s^{+}\gamma$ decays. 
To this, we assign the observed scalar nonet, $\sigma(600)$, 
$f_0(980)$, $a_0(980)$~\cite{PDG04}and $\kappa(800)$~\cite{E791},  
to the scalar four-quark $[qq][\bar q\bar q]$ mesons as suggested long 
time ago~\cite{Jaffe} and as supported by many analyses, for example, in 
Refs.~\cite{Achasov,Bugg,Maiani,CT,Suganuma}, and adopt the observed rates 
%%%%%%%%%%%%%%%%%%%%%%%%%%%%%%%%%%%%%%%%%%%%%%%%%%%%%%%%%%%%%%%%%%%%%%%%%
$\Gamma(a_0(980)\rightarrow \eta\pi)_{\rm exp}$ and 
$\Gamma(\phi\rightarrow a_0(980)\gamma)_{\rm exp}$ 
%%%%%%%%%%%%%%%%%%%%%%%%%%%%%%%%%%%%%%%%%%%%%%%%%%%%%%%%%%%%%%%%%%%%%%%%%
in Ref.~\cite{PDG04} as the input data~\cite{HT-isospin}. In this way, 
we see later that the experimental data on the $D_{s}^{*+}\gamma$ and 
$D_s^+\pi^0$ decays favor the assignment of $D_{s0}^+(2317)$ to the 
$I_3=0$ component of iso-triplet scalar four-quark mesons with charm and 
strangeness. It may be considered as an evidence for existence of scalar 
four-quark mesons with charm. 

In the next section, we will give a very brief review on four-quark
mesons with light flavors and extend them to the ones with charm. 
Then, we will study why $D_{s0}^+(2317)$ is so narrow. It is predicted 
that four-quark $[cq][\bar q\bar q]$ mesons are narrow or rather stable 
in {\bf III}. In {\bf IV} and {\bf V},  radiative decays and isospin 
non-conserving decays of $D_{s0}^+(2317)$ will be studied, 
respectively. The results will be compared with the
experimental constraint, Eq.~(\ref{eq:CLEO}). The broad bump 
in the $D\pi$ mass distribution observed by the BELLE and by the FOCUS 
will be discussed in {\bf VI}. A brief summary will be given in the final 
section. 

%%%%%%%%%%%%%%%%%%%%%%%%%%%%%%%%%
\section{Four-quark mesons}
%%%%%%%%%%%%%%%%%%%%%%%%%%%%%%%%%

Before studying four-quark mesons, we review very briefly potentials, 
%%%%%%%%%%%%%%%%%%%%%%%%%%%%%%%%%%%%%%%%%%%%%%%%%%%%%%%%%%%%%%%%%%%%%%%%%%
\begin{equation}
V_{qq}({\bf r})=\sum\Lambda_i\Lambda_i v({\bf r}) \quad{\rm and}\quad 
V_{q\bar q}({\bf r})=-\sum\Lambda_i\Lambda_i v({\bf r}), 
\end{equation}
%%%%%%%%%%%%%%%%%%%%%%%%%%%%%%%%%%%%%%%%%%%%%%%%%%%%%%%%%%%%%%%%%%%%%%%%%%
between two quarks and between a quark and an anti-quark, respectively, 
mediated by a vector meson with an extra $SU(3)$ degree of 
freedom~\cite{Hori} corresponding to the "color". The results are 
summarized in Table~2, because they are still instructive although they 
have been studied much earlier than the discovery of the 
color~\cite{color}. 

As seen in Table~2, the force between two quarks (or between two 
anti-quarks) is attractive when they are of ${\bf \bar 3_c}$ 
(or ${\bf 3_c}$) but repulsive when they are of ${\bf 6_c}$ 
(or ${\bf \bar 6_c}$) while the force between a quark and 
an anti-quark is attractive and much stronger when they are of color singlet. 
%%%%%%%%%%%%%%%%%%%%%%%%%%%%%%%%%%%%%%%%%%%%%%%%%%%%%%%%%%%%%%%%%%%%%%%%%%
%\vspace{2mm}
%\newpage
%%%%%%%%%%%%%%%%%%%
\begin{center}
\begin{quote}
{Table~2. 
Potentials mediated by a vector meson with $SU(3)$ "color". 
}
\end{quote}\vspace{3mm}

\begin{tabular}
{| c | c | c | c | c |}
\hline
&\multicolumn{2}{|c|}
{$qq$}
&\multicolumn{2}{|c|}{$q\bar q$}
\\
\hline
 & & & & \vspace{-4mm}\\
$SU_c(3)$
&
$\bf{\bar 3_c}$
&
$\bf{6_c}$
&
$\bf{8_c}$
&
$\bf{1_c}$
\\
 & & & & \vspace{-5mm}\\
\hline
 & & & & \vspace{-4mm}\\
\vspace{0mm}
Potential
&$\displaystyle{-{8\over 3}\langle{v}\rangle}$
&\,\,\,$\displaystyle{4\over 3}\langle{v}\rangle$
& \,\,\,$\displaystyle{{2\over 3}\langle{v}\rangle}$ 
& $\displaystyle{-{16\over 3}\langle{v}\rangle}$
\vspace{-4mm}
\\
 & & & & \vspace{-0mm}\\
\hline
\end{tabular}

\end{center}
\vspace{2mm}
%%%%%%%%%%%%%%%%%%%%%%%%%%%%%%%%%%%%%%%%%%%%%%%%%%%%%%%%%%%%%%%%%%%%%%%%%%%

Now, four-quark meson states can be classified into the following four 
groups~\cite{Jaffe},
%%%%%%%%%%%%%%%%%%%%%%%%%%%%%%%%%%%%%%%%%%%%%%%%%%%%%%%%%%%%%%%%%%%%%%%%
\begin{equation}
\{qq\bar q\bar q\} = 
[qq][\bar q\bar q] \oplus (qq)(\bar q\bar q) 
\oplus \{[qq](\bar q\bar q)\pm (qq)[\bar q\bar q]\}
                                                 \label{eq:4-quark}
\end{equation}
%%%%%%%%%%%%%%%%%%%%%%%%%%%%%%%%%%%%%%%%%%%%%%%%%%%%%%%%%%%%%%%%%%%%%%%%
where $(\,)$ and $[\,]$ denote symmetry and anti-symmetry, respectively, 
under the exchange of flavors between them. The first two on the 
right-hand-side (r.h.s.) of Eq.~(\ref{eq:4-quark}) can have 
$J^{P(C)}=0^{+(+)}$ while the last two have $J^{P}=1^{+}$. 
We are now interested only in scalar mesons, so that we consider the first 
two. Each of them is again classified into two classes since there are 
two different ways to produce color singlet $\{qq\bar q\bar q\}$ states, 
i.e., two different combinations of color 
${\bf \bar 3_c}\times{\bf 3_c}$ and ${\bf 6_c}\times {\bf \bar 6_c}$ 
representations to produce totally colorless states. 
Because of the property of the forces between two quarks 
(or anti-quarks) discussed before, it is expected that the scalar 
$[qq][\bar q\bar q]$ mesons of ${\bf \bar 3_c}\times {\bf 3_c}$ of 
$SU_c(3)$ can be the lowest lying four-quark mesons. However, these two 
states of  
%%%%%%%%%%%%%%%%%%%%%%%%%%%%%%%%%%%%%%%%%%%%%%%%%%%%%%%%%%%%%%%%%%%%%%%
${\bf \bar 3_c}\times {\bf 3_c}$ and ${\bf 6_c}\times {\bf \bar 6_c}$,  
%%%%%%%%%%%%%%%%%%%%%%%%%%%%%%%%%%%%%%%%%%%%%%%%%%%%%%%%%%%%%%%%%%%%%%%
in general, can mix with each other. Nevertheless, we neglect hereafter 
the small mixing of ${\bf 6_c}\times{\bf \bar 6_c}$ in the lighter class 
of $[qq][\bar q\bar q]$ mesons for simplicity. The MIT bag model with 
the bag potential and a spin-spin force arising from one gluon exchange 
shows that the $[qq][\bar q\bar q]$ mesons which are dominantly of 
${\bf \bar 3_c}\times {\bf 3_c}$ and make a ${\bf 9_f}$-plet of the
flavor $SU_f(3)$ are the lowest lying states~\cite{Jaffe}. 
The wave functions of the $[qq][\bar q\bar q]$ mesons with respect to 
flavors and their mass values which have been calculated in 
Ref.~\cite{Jaffe} are listed in Table~3. 
As seen in the table, the mass relations among the $[qq][\bar q\bar q]$ 
nonet members seem to be well realized by the observed candidates, 
although the calculated mass values are systematically a little larger 
than the observed ones. 
In this way, we can easily understand the approximate degeneracy between 
$f_0(980)$ and $a_0(980)$, and the mass hierarchy in the nonet. 
%%%%%%%%%%%%%%%%%%%%%%%%%%%%%%%%%%%%%%%%%%%%%%%%%%%%%%%%%%%%%%%%%%%%%%%%%%%
%\newpage
\vspace{0mm}
%%%%%%%%%%%%%%%%%%%
\begin{center}
\begin{quote}
{Table~3. Ideally mixed scalar $[qq][\bar q\bar q]$ mesons 
(with $q=u,\,d,\,s$) and their candidates, where 
$S$ and $I$ denote strangeness and isospin quantum numbers. 
The calculated mass values have been taken from Ref.~\cite{Jaffe}. 
}
\end{quote}\vspace{3mm}

\begin{tabular}
{|c|c|c|c|c|c|}
\hline
$\,\, S\,\,$
&$\,\, I=1\,\,$
&$\,\,I={1\over 2}\,\,$
&$\,\, I=0\,\,$
&$\,\,$Mass(GeV)$\,\,$
&Candidate
\\
\hline
$1$
&
&$\hat\kappa\sim[ud][\bar n\bar s]$
&
&  
0.90  
&  
$\kappa(800)$   
\\
\hline
0
&  \begin{tabular}{c}
$[ns][\bar n\bar s]_{I=1}\sim$
{$\hat\delta^s$}\\
\\
\end{tabular}
&
&   \begin{tabular}{l}
$\hat\sigma^s\sim[ns][\bar n\bar s]_{I=0}$
\\
$\hat\sigma\sim[ud][\bar u\bar d]\hspace{5mm}\hspace{0mm}$
\end{tabular}
&  \begin{tabular}{c}
{1.10}\\
{0.65}\\
\end{tabular}
&\begin{tabular}{c}
$a_0(980)$, $f_0(980)$\\
$\sigma(600)$   
\end{tabular}
\\
\hline 
\end{tabular}\vspace{1mm}
\end{center}
%%%%%%%%%%%%%%%%%%%%%%%%%%%%%%%%%%%%%%%%%%%%%%%%%%%%%%%%%%%%%%%%%%%%%%%%%%

The $(qq)(\bar q\bar q)$ mesons also can have $J^{P(C)}=0^{+(+)}$. 
However, their masses which have been estimated in the same way as the 
above are much higher than those of the $[qq][\bar q\bar q]$ 
mesons~\cite{Jaffe}, so that we do not consider the $(qq)(\bar q\bar q)$ 
any more in this article. 
%%%%%%%%%%%%%%%%%%%%%%%%%%%%%%%%%%%%%%%%%%%%%%%%%%
%\section{Why is ${\bf D_{s0}^+(2317)}$ so narrow ?}
%\section{Narrow width of ${\bf D_{s0}^+(2317)}$}
%%%%%%%%%%%%%%%%%%%%%%%%%%%%%%%%%%%%%%%%%%%%%%%%%%
%%%%%%%%%%%%%%%%%%%%%%%%%%%%%%%%%%%%%%%%%%%%%%%%%%%%%%%%%%%%%%%%%%%%%%%%%%
%\newpage
\vspace{0mm}
%%%%%%%%%%%%%%%%%%%%%%
\begin{center}
\begin{quote}
{Table~4. Ideally mixed scalar $[cq][\bar q\bar q]$ mesons 
(with $q=u,\,d,\,s)$, where $C$, $S$ and $I$ denote charm, strangeness 
and isospin quantum numbers. Their mass values 
are estimated in the text. 
}
\end{quote}
\vspace{3mm}

\begin{tabular}
{|c|c|c|c|c|c|}
\hline
$\,\, C\,\,$&$\,\, S\,\,$
&$\,\, I=1\,\,$
&$\,\,I={1\over 2}\,\,$
&$\,\, I=0\,\,$
&$\,\,$Mass(GeV)$\,\,$
\\
\hline
&$1$
&
{$\hat F_I^{0,+,++}\sim [cn][\bar s\bar n]_{I=1}$} &
&
{$\hat F_0^+\sim [cn][\bar s\bar n]_{I=0}$}  &
2.32($\ddagger$)  \\
\cline{2-6}
\begin{tabular}{c}
1\\
\end{tabular}
&0
&
&
\begin{tabular}{c}
{$\hat D^{s+,0}\sim [cs][\bar s\bar n]$}\\
{$\hat D^{+,0}\sim [cn][\bar u\bar d]$}   
\end{tabular}
&
&\begin{tabular}{c}
\hspace{-4mm}{2.42}\\
\hspace{-4mm}{2.22}  
\end{tabular}
\\
\cline{2-6} 
&-1
&
&
&
{$\hat E^0\sim [cs][\bar u\bar d]$}  &
\hspace{-4mm}{2.32}    \\
\hline
\end{tabular}\vspace{0mm}
\end{center}
%%%%%%%%%%%%%%%%%%%%%%%%%%%%%%%%%%%%%%%%%%%%%%%%%%%%%%%%%%%%%%%%%%%%%%%%%%%
\noindent{\hspace{105mm}($\ddagger$) Input data} 
\vspace{0mm}\\
%%%%%%%%%%%%%%%%%%%%%%%%%%%%%%%%%%%%%%%%%%%%%%%%%%%%%%%%%%%%%%%%%%%%%%%%%%%

Extension to charm scalar four-quark mesons is 
straightforward~\cite{Terasaki-D_s}, i.e., to replace a light quark $q$ 
in the $[qq][\bar q\bar q]$ by a charm quark $c$. 
Open charm scalar four-quark $[cq][\bar q\bar q]$ mesons are listed in 
Table 4. However, their mass values are not yet definite, although 
there have been various efforts~\cite{MPPR,QCDSR-4q,KO,Bicudo} to 
estimate their masses.  Therefore, we list tentative results from 
a simple quark counting with 
%%%%%%%%%%%%%%%%%%%%%%%%%%%%%%%%%%%%%%%%%%%%%%%%%%%%%%%%%%%%%%%%%%%%%%%%%%
$\Delta_s = m_s - m_n \simeq m_{D_s} - m_{D} \simeq 0.10$ GeV   
%%%%%%%%%%%%%%%%%%%%%%%%%%%%%%%%%%%%%%%%%%%%%%%%%%%%%%%%%%%%%%%%%%%%%%%%%%
around 2 GeV scale, 
where we have assigned $D_{s0}^+(2317)$ to the 
iso-triplet $\hat F_I^+$ meson and have taken $m_{D_{s0}} = 2.32$ GeV as 
the input data. 

Next, we study isospin conserving decays of scalar $[cq][\bar q\bar q]$ 
mesons, assigning $D_{s0}^+(2317)$ to the iso-triplet scalar 
$\hat F_I^+\sim [cn][\bar s\bar n]_{I=1}$ meson and comparing them with 
the observed decay $a_0(980)\rightarrow \eta\pi$, where the observed 
$\sigma(600)$, $\kappa(800)$, $a_0(980)$ and $f_0(980)$ are assigned 
to the scalar $[qq][\bar q\bar q]$ mesons, $\hat\sigma$, $\hat\kappa$, 
$\hat\delta^s$ and $\hat\sigma^s$, respectively, as discussed before. 
In this way, we see, below, why $\hat F_I^+$ is so narrow, and the other 
decays will be studied in the later sections. 

To this, we write the rate for the 
$A({\bf p})\rightarrow B({\bf p'})\pi({\bf q})$ decay as 
%%%%%%%%%%%%%%%%%%%%%%%%%%%%%%%%%%%%%%%%%%%%%%%%%%%%%%%%%%%%%%%%%%%%%%%%%%
\begin{equation}
\Gamma(A\rightarrow B\pi)
=\Biggl({1\over 2J_A+1}\Biggr)\Biggl({q_c\over 8\pi m_A^2}\Biggr)
\sum_{\rm spins}|M(A\rightarrow B\pi)|^2,
                                        \label{eq:rate-general}
\end{equation}
%%%%%%%%%%%%%%%%%%%%%%%%%%%%%%%%%%%%%%%%%%%%%%%%%%%%%%%%%%%%%%%%%%%%%%%%%%
where $J_A$, $q_c$ and $M(A\rightarrow B\pi)$ are the spin of 
the parent particle $A$, the momentum of the final particles in 
the rest frame of the parent $A$, and the amplitude for the decay, 
respectively. To calculate the amplitude, we use the PCAC 
(partially conserved axial vector current) hypothesis and 
a hard pion approximation in the infinite momentum frame (IMF), 
i.e., $|{\bf p}|\rightarrow \infty$~\cite{hard-pion,suppl}. 
It is an innovation of the old current algebra. 
In this approximation, the amplitude is evaluated at a little 
unphysical point, i.e., $m_\pi^2\rightarrow 0$, 
%%%%%%%%%%%%%%%%%%%%%%%%%%%%%%%%%%%%%%%%%%%%%%%%%%%%%%%%%%%%%%%%%%%%%%%%%%
\begin{equation}
M(A\rightarrow B\pi) \simeq \Biggl({m_A^2 - m_B^2\over f_\pi}
\Biggr)\langle{B|A_\pi|A}\rangle,   
                                                \label{eq:hard-pion}
\end{equation}
%%%%%%%%%%%%%%%%%%%%%%%%%%%%%%%%%%%%%%%%%%%%%%%%%%%%%%%%%%%%%%%%%%%%%%%%%%
where $A_\pi$ is the axial counterpart of the isospin. 
The {\it asymptotic matrix element} of $A_\pi$ (matrix element 
of $A_\pi$ taken between single hadron states with infinite 
momentum), $\langle{B|A_\pi|A}\rangle$, gives the dimensionless 
coupling strength. 

The $\hat F_I^+\rightarrow D_s^+\pi^0$ decay is only one decay of 
$\hat F_I^+$ which conserves isospin, so that the narrow width of 
$\hat F_I^+$ means that the rate for the decay is small, i.e., 
overlap of wavefunctions between the final $D_s^+\pi^0$ and 
the initial $\hat F_I^+$ is small. In the present case, it corresponds 
to a small value of $|\langle{D_s^+|A_{\pi^0}|\hat F_I^+}\rangle|$. 
Such a small overlap of wavefunctions is possible in decays 
of heavy (charm) four-quark mesons into two pseudoscalar 
mesons~\cite{Terasaki-yitpws,Terasaki-hadron03,Terasaki-mquark} as seen 
below. The scalar four-quark $[qq][\bar q\bar q]$ meson state can be 
decomposed into a sum 
of products of two $\{q\bar q\}$ states with various color and spin; 
%%%%%%%%%%%%%%%%%%%%%%%%%%%%%%%%%%%%%%%%%%%%%%%%%%%%%%%%%%%%%%%%%%%%%%%%%%
\begin{eqnarray}
|[qq]^{\bf 1_s}_{\bf \bar 3_c}
[\bar q\bar q]^{\bf 1_s}_{\bf 3_c}\rangle_{\bf 1_c}^{\bf 1_s} 
&&{=}\quad { -\sqrt{1\over 4}}
{\times} {\sqrt{1\over 3}}
{|\{q\bar q\}^{\bf 1_s}_{\bf 1_c}
\{q\bar q\}^{\bf 1_s}_{\bf 1_c}\rangle_{\bf 1_c}^{\bf 1_s}}
\quad{ +} \quad
{\sqrt{3\over 4}}  {\times}
{\sqrt{1\over 3}}
|\{q\bar q\}^{\bf 3_s}_{\bf 1_c}
\{q\bar q\}^{\bf 3_s}_{\bf 1_c}\rangle_{\bf 1_c}^{\bf 1_s} \nonumber\\
&&\hspace{3mm}
\quad - { \sqrt{1\over 4}}
{\times} {\sqrt{2\over 3}}
{|\{q\bar q\}^{\bf 1_s}_{\bf 8_c}
\{q\bar q\}^{\bf 1_s}_{\bf 8_c}\rangle_{\bf 1_c}^{\bf 1_s}}
\quad{ +} \quad
{\sqrt{3\over 4}}  {\times}
{\sqrt{2\over 3}}
|\{q\bar q\}^{\bf 3_s}_{\bf 8_c}
\{q\bar q\}^{\bf 3_s}_{\bf 8_c}\rangle_{\bf 1_c}^{\bf 1_s}, 
                                            \label{eq:overlapping}
\end{eqnarray}
%%%%%%%%%%%%%%%%%%%%%%%%%%%%%%%%%%%%%%%%%%%%%%%%%%%%%%%%%%%%%%%%%%%%%%%%%%
where the first and second coefficients of each term on the r.h.s. of 
the above equation are given by the crossing matrices for spin and 
color~\cite{Jaffe}, respectively. The superscripts $\bf{1_s}$ and 
$\bf{3_s}$ denote the spin singlet and triplet, and the subscripts  
$\bf{1_c}$ and $\bf{8_c}$ the color singlet and octet. Since the 
$\{q\bar q\}$ pairs in the second line of the r.h.s. are of ${\bf 8_c}$, 
they are unable to go into a final state of two 
physical mesons, unless a gluon is exchanged between them. When a gluon 
is exchanged, however, the states in the second line can supply the 
states with the configuration of color and spin in the first line, i.e., 
the configuration of color and spin given in Eq.~(\ref{eq:overlapping}) 
will be reshuffled by a gluon exchange, and then the states in the 
second line can go to two pseudoscalar and two vector meson states. 

Such a gluon exchange will be more effective at lower energy scale 
while less at higher energy scale. Since it is known that the $s$-quark 
is considerably "slim" at the 2 GeV scale~\cite{Gupta}, gluon couplings 
around the scale of charm meson mass will be rather perturbative and 
therefore it is expected that gluon exchanges between the two 
$\{q\bar q\}_{\bf 8_c}$'s on the r.h.s. will be less effective in the 
charmed four-quark mesons. In contrast, in the case of light four-quark 
mesons, they will be non-perturbative and the gluon exchange between 
the two $\{q\bar q\}_{\bf 8_c}$'s  will be highly effective, so that 
the configurations of color and spin can be very easily reshuffled. 

Now we compare the $\hat F_I^+\rightarrow D_s^+\pi^0$ with the 
experimentally known $\hat \delta^{s+}\rightarrow \eta\pi^+$. 
The $\eta$ can be decomposed into a sum of $n\bar n$ and $s\bar s$ 
components ($\eta^n$ and $\eta^s$, respectively) as 
%%%%%%%%%%%%%%%%%%%%%%%%%%%%%%%%%%%%%%%%%%%%%%%%%%%%%%%%%%%%%%%%%%%%%%%%%
\begin{equation}
\eta = \cos\Theta\cdot \eta^n - \sin\Theta\cdot \eta^s 
\end{equation}
%%%%%%%%%%%%%%%%%%%%%%%%%%%%%%%%%%%%%%%%%%%%%%%%%%%%%%%%%%%%%%%%%%%%%%%%%
where $\Theta = \chi + \theta_P$ with $\cos\chi=\sqrt{1/3}$ and 
the usual $\eta$-$\eta'$ mixing angle 
$\theta_P\simeq -20^\circ$~\cite{PDG04}. 
To see difference of wavefunction overlap between the above two decays, 
we introduce a parameter given by a ratio of the asymptotic matrix 
elements, 
%%%%%%%%%%%%%%%%%%%%%%%%%%%%%%%%%%%%%%%%%%%%%%%%%%%%%%%%%%%%%%%%%%%%%%%%%
\begin{equation}
\beta \equiv {\sqrt{2}\langle{D_s^+|A_{\pi^0}|\hat F_I^+}\rangle
\over \langle{\eta^s|A_{\pi^-}|\hat \delta^{s+}}\rangle}.
                                                       \label{eq:beta}
\end{equation}
%%%%%%%%%%%%%%%%%%%%%%%%%%%%%%%%%%%%%%%%%%%%%%%%%%%%%%%%%%%%%%%%%%%%%%%%
A naive asymptotic $SU_f(4)$ symmetry~\cite{afs} in which different 
configurations of color and spin are not cared and overlap of spatial 
wavefunctions is assumed to be in the symmetry limit implies that 
$\beta = 1$. However, the fact that a four-quark $[qq][\bar q\bar q]$ 
state can be decomposed into a sum of products of two $\{q\bar q\}$ 
pairs with various configurations of color and spin is very important 
in decays of four-quark mesons as seen above. For simplicity, we here 
consider the limiting case that no reshuffling in the 
$\hat F_I^+\rightarrow D_s^+\pi^0$ (at higher energy scale) 
but the full reshuffling in the $\hat\delta^{s+}\rightarrow \eta\pi^+$ 
decay (at lower energy scale). In this limit, the overlapping factor 
$\beta$ with respect to the color and spin wavefunctions could be given 
by 
%%%%%%%%%%%%%%%%%%%%%%%%%%%%%%%%%%%%%%%%%%%%%%%%%%%%%%%%%%%%%%%%%%%%%%%%%
\begin{equation}
|\beta|_{SU_f(4)}^2 ={1\over 3}\times{1\over 4} ={1\over 12},    
                                              \label{eq:beta-square}
\end{equation}
%%%%%%%%%%%%%%%%%%%%%%%%%%%%%%%%%%%%%%%%%%%%%%%%%%%%%%%%%%%%%%%%%%%%%%%%
when the overlap of the spatial wavefunctions is in the $SU_f(4)$ 
symmetry limit.  

Taking tentatively  
%%%%%%%%%%%%%%%%%%%%%%%%%%%%%%%%%%%%%%%%%%%%%%%%%%%%%%%%%%%%%%%%%%%%%
$\Gamma(\hat\delta^{s+}\rightarrow\eta\pi^+)\sim 70$ MeV 
%%%%%%%%%%%%%%%%%%%%%%%%%%%%%%%%%%%%%%%%%%%%%%%%%%%%%%%%%%%%%%%%%%%%%
from the experimental data~\cite{PDG04} that 
%%%%%%%%%%%%%%%%%%%%%%%%%%%%%%%%%%%%%%%%%%%%%%%%%%%%%%%%%%%%%%%%%%%%%%
$\Gamma_{a_0(980)}=50 - 100$ MeV and 
$a_0(980)\rightarrow\eta\pi$ dominates the decays of $a_0(980)$  
%%%%%%%%%%%%%%%%%%%%%%%%%%%%%%%%%%%%%%%%%%%%%%%%%%%%%%%%%%%%%%%%%%%%%%
and using Eq.~(\ref{eq:rate-general}) with Eq.~(\ref{eq:hard-pion}),  
we estimate 
%%%%%%%%%%%%%%%%%%%%%%%%%%%%%%%%%%%%%%%%%%%%%%%%%%%%%%%%%%%%%%%%%%%%%%
\begin{equation}
|\langle{\eta^s|A_{\pi^-}|\hat\delta^{s+}}\rangle| \sim 0.80  
                                        \label{eq:4-quark-coupling}
\end{equation}
%%%%%%%%%%%%%%%%%%%%%%%%%%%%%%%%%%%%%%%%%%%%%%%%%%%%%%%%%%%%%%%%%%%%%%%
The above value of 
$|\langle{\eta^s|A_{\pi^-}|\hat\delta^{s+}}\rangle|$ and 
Eq.~(\ref{eq:beta-square}) lead us to 
%%%%%%%%%%%%%%%%%%%%%%%%%%%%%%%%%%%%%%%%%%%%%%%%%%%%%%%%%%%%%%%%%%%%%%
\begin{equation}
\Gamma(\hat F_I^+\rightarrow D_s^+\pi^0)_{SU_f(4)} 
                \sim 9\,\,{\rm MeV},      \label{eq:rate-of-F_I-SU(4)}
\end{equation}
%%%%%%%%%%%%%%%%%%%%%%%%%%%%%%%%%%%%%%%%%%%%%%%%%%%%%%%%%%%%%%%%%%%%%%%
which dominates the width of $\hat F_I^+$. The above value of the rate 
is a little larger than the experimental upper bound of width of 
$D_{s0}^+(2317)$. However, it is not serious because deviation from 
the $SU_f(4)$ symmetry limit of spatial wavefunction overlap is not 
considered at this stage. $SU_f(4)$ symmetry of spatial wavefunction 
overlap overestimates asymptotic matrix elements of charges, ususally 
by about $20 - 30$ \%, as seen below. 

A measure of the (asymptotic) flavor symmetry breaking (in overlap of 
spatial wavefunctions) will be seen in deviations from unity of values of 
form factor, $f_+(0)$, of related vector current at zero 
momentum transfer squared. The estimated values of $f_+(0)$ are 
%%%%%%%%%%%%%%%%%%%%%%%%%%%%%%%%%%%%%%%%%%%%%%%%%%%%%%%%%%%%%%%%%%%%
\begin{eqnarray}
f_+^{(\pi K)}(0)&&= 0.961 \pm 0.008, \label{eq:LR}\\
f_+^{(\bar K D)}(0)&&= 0.74 \pm 0.03,  \label{eq:PDG96}\\
\left[f_+^{(\pi D)}(0)\right]\left/
\left[f_+^{(\bar K D)}(0)\right]
\right.
&&= 1.00 \pm 0.11 \pm 0.02, \label{eq:E687} \\
&&= 0.99\pm 0.08,  \label{eq:CLEO97}
\end{eqnarray}
%%%%%%%%%%%%%%%%%%%%%%%%%%%%%%%%%%%%%%%%%%%%%%%%%%%%%%%%%%%%%%%%%%%%
where the values in Eqs.~(\ref{eq:LR}) -- (\ref{eq:CLEO97}) have 
been taken from Refs.~\cite{LR} --  ~\cite{CLEO97}, respectively. 
They imply that the asymptotic flavor $SU_f(3)$ symmetry works 
well while the $SU_f(4)$ overestimates by about 20 -- 30 $\%$. 
This statement is confirmed by the observation that the asymptotic 
symmetry has predicted the rates~\cite{suppl,HOS}, 
%%%%%%%%%%%%%%%%%%%%%%%%%%%%%%%%%%%%%%%%%%%%%%%%%%%%%%%%%%%%%%%%%%%%
$\Gamma(D^{*+}\rightarrow D^0\pi^+)_{SU_f(4)} \simeq 96$ keV and 
$\Gamma(D^{*+}\rightarrow D^+\pi^0)_{SU_f(4)} \simeq 42$ keV, 
%%%%%%%%%%%%%%%%%%%%%%%%%%%%%%%%%%%%%%%%%%%%%%%%%%%%%%%%%%%%%%%%%%%%
which are larger by about $50\,\,\%$ than the observed values, 
%%%%%%%%%%%%%%%%%%%%%%%%%%%%%%%%%%%%%%%%%%%%%%%%%%%%%%%%%%%%%%%%%%%%
$\Gamma(D^{*+}\rightarrow D^0\pi^+)_{\rm exp} = 65 \pm 18$ keV 
and 
$\Gamma(D^{*+}\rightarrow D^+\pi^0)_{\rm exp} = 30 \pm 8$ keV,  
%%%%%%%%%%%%%%%%%%%%%%%%%%%%%%%%%%%%%%%%%%%%%%%%%%%%%%%%%%%%%%%%%%%%
obtained from the measured decay width~\cite{CLEO-D^*},
%%%%%%%%%%%%%%%%%%%%%%%%%%%%%%%%%%%%%%%%%%%%%%%%%%%%%%%%%%%%%%%%%%%%
$\Gamma_{D^{*\pm}}=96 \pm 4 \pm 22$ keV, 
%%%%%%%%%%%%%%%%%%%%%%%%%%%%%%%%%%%%%%%%%%%%%%%%%%%%%%%%%%%%%%%%%%%%
and the branching fractions compiled in Ref.~\cite{PDG04}. 
The above suggests that asymptotic $SU_f(4)$ symmetry 
overestimates the rate for the $D^*\rightarrow D\pi$ decays, by 
about 50 \%,  as expected from the above values of the form factor 
$f_+(0)$. However, for simplicity, we will use 
asymptotic $SU_f(4)$ symmetry relations among asymptotic matrix
elements of $A_\pi$ and $A_K$ in our estimates of decay rates. 
When we take account for the symmetry breaking, we will note it. 

The above asymptotic $SU_f(4)$ symmetry breaking reduces the result in 
Eq.~(\ref{eq:rate-of-F_I-SU(4)}) by about 40 $-$ 60 \% 
because the amplitude is proportional to the asymptotic matrix element 
of axial charge, i.e., 
%%%%%%%%%%%%%%%%%%%%%%%%%%%%%%%%%%%%%%%%%%%%%%%%%%%%%%%%%%%%%%%%%%%%%%
\begin{equation}
\Gamma(\hat F_I^+\rightarrow D_s^+\pi^0) \sim 4 - 5\,\,{\rm MeV}. 
                                               \label{eq:rate-of-F_I}
\end{equation}
%%%%%%%%%%%%%%%%%%%%%%%%%%%%%%%%%%%%%%%%%%%%%%%%%%%%%%%%%%%%%%%%%%%%%%%
It is consistent with the experimental data on the width of 
$D_{s0}^+(2317)$ in Eq.~(\ref{eq:D_s-data}). It seems to imply that 
the above limiting situation providing $|\beta|_{SU_f(4)}^2 = 1/12$ is 
approximately realized and that the asymptotic $SU_f(4)$ symmetry 
breaking arising from the overlap of spatial wavefunctions including 
a four-quark state is not very far from the one between the conventional 
meson states. 

%%%%%%%%%%%%%%%%%%%%%%%%%%%%%%%%%%%%%%%%%%
\section{Isospin Conserving Decays}
%%%%%%%%%%%%%%%%%%%%%%%%%%%%%%%%%%%%%%%%%%

In this section, we study isospin conserving two-body decays of the scalar 
$[cq][\bar q\bar q]$ mesons. $SU_f(3)$ relations of asymptotic matrix 
elements of $A_{\pi}$ are given by 
%%%%%%%%%%%%%%%%%%%%%%%%%%%%%%%%%%%%%%%%%%%%%%
\begin{eqnarray}
&& 
\langle{D_s^+|A_{\pi^-}|\hat F_I^{++}}\rangle
=\sqrt{2}\langle{D_s^+|A_{\pi^0}|\hat F_I^{+}}\rangle
=\langle{D_s^+|A_{\pi^+}|\hat F_I^{0}}\rangle 
\nonumber\\
&&  
=-\langle{D^0|A_{\pi^-}|\hat D^{+}}\rangle
=2\langle{D^+|A_{\pi^0}|\hat D^{+}}\rangle
=-2\langle{D^0|A_{\pi^0}|\hat D^{0}}\rangle    
=-\langle{D^+|A_{\pi^+}|\hat D^0}\rangle .     
                                               \label{eq:su_f(3)}
\end{eqnarray}
%%%%%%%%%%%%%%%%%%%%%%%%%%%%%%%%%%%%%%%%%%%%%%
Inserting Eq.~(\ref{eq:hard-pion}) with Eq.~(\ref{eq:su_f(3)}) into 
Eq.~(\ref{eq:rate-general}) and using the width, 
%%%%%%%%%%%%%%%%%%%%%%%%%%%%%%%%%%%%%%%%%%%%%%%%%%%%%%%%%%%%%%%%%%%%%%
$\Gamma(\hat F_I^+) \simeq 
\Gamma(\hat F_I^+\rightarrow D_s^+\pi^0) \sim 4.5\,\,{\rm MeV}$,  
%%%%%%%%%%%%%%%%%%%%%%%%%%%%%%%%%%%%%%%%%%%%%%%%%%%%%%%%%%%%%%%%%%%%%%%
from Eq.~(\ref{eq:rate-of-F_I}) as the input data, we obtain 
the results in Table~5. As seen in the table, the iso-triplet $\hat F_I$ 
mesons have the same width as the input. The iso-doublet $\hat D$ mesons 
are broader by about 50 \% than the input data because they have two 
possible decay modes through isospin conserving strong interactions. 
Another iso-doublet $\hat D^s$ mesons will be just on the threshold of 
$D\eta$ decay, if the true mass of $\hat D^s$ is close to the estimated 
one. It is only one strong decay which is kinematically allowed.  
Since its rate is sensitive to the mass of $\hat D^s$ whose 
precise value has not been known, however, we here do not calculate it. 
Nevertheless, it is expected that its width is much narrower than the 
input. The iso-singlet $\hat F_0^+$ meson will decay dominantly into 
$D_s^{*+}\gamma$~\cite{HT-isospin} as will be studied later, 
if its mass is really lower than the $DK$ threshold. 
The exotic $\hat E^0$ would decay through weak interactions unless its 
mass is higher than the $D\bar K$ threshold~\cite{Lipkin,ST}. 
Therefore, all the scalar $[cq][\bar q\bar q]$ mesons will be observed 
as narrow resonances except for $\hat E^0$ which will be observed as a 
rather stable meson unless it is heavier than the $D\bar K$ threshold. 
Even if it is more massive than $m_D\,+\,m_K$, it would be as narrow 
as the other members of the scalar $[cq][\bar q\bar q]$ mesons. 
%%%%%%%%%%%%%%%%%%%%%%%%%%%%%%%%%%%%%%%%%%%%%%%%%%%%%%%%%%%%%%%%%%%%%%%
\vspace{2mm}
%\newpage
%%%%%%%%%%%%%%%%%%%%%%
\begin{center}
\begin{quote}
Table~5. 
Dominant decays of scalar {$[cq][\bar q\bar q]$} mesons. The input data, 
{$\Gamma(\hat F_I^+)\simeq \Gamma(\hat F_I^+\rightarrow D_s^+\pi^0)
\sim 4.5\,\,{\rm MeV}$}, 
is taken from the estimated width, Eq.~(\ref{eq:rate-of-F_I}), in the text. 
\end{quote}  \vspace{3mm}

\begin{tabular}{l l l}
\hline
\quad Parent(Mass in GeV)\quad & \quad Final State\quad 
& \quad Decay Rate (in MeV)\quad
\\
\hline 
\,\,\,\,\begin{tabular}{l}
{$\hat F_I^{++}(2.32)$}\\             
{$\hat F_I^+(2.32)$}\\
{$\hat F_I^0(2.32)$}
\end{tabular}\quad 
&\quad 
\begin{tabular}{l}
{$D_s^+\pi^+$}\\
{$D_s^+\pi^0$}\\
{$D_s^+\pi^-$}
\end{tabular}\quad 
&
$\left.\begin{tabular}{l}
\\
\\
\end{tabular}
\right\}$
{$\sim 4.5$}\\
\hline
 \quad $\hat D^+(2.22)$ \quad
&\quad \begin{tabular}{l}
 {$D^0\pi^+$}\\
 {$D^+\pi^0$}
\end{tabular}\quad
&\quad\begin{tabular}{l}
{$\sim 4.5$}\\
{$\sim 2.3$}
\end{tabular}\quad\\
\quad {$\hat D^0(2.22)$}\quad 
& \quad\begin{tabular}{l}
 {$D^+\pi^-$}\\
 {$D^0\pi^0$}
\end{tabular}\quad
&\quad\begin{tabular}{l}
 {$\sim 4.5$}\\
{$\sim 2.3$}
\end{tabular}\quad \\
\hline 
\quad {$\hat D^s(2.42)$}\quad 
&\quad\hspace{0.5mm}{$D\eta$} (or  {$D^{*+}\gamma$})\quad
& \quad( {$\ll 4.5$})\quad\\
\hline 
\quad {$\hat F_0^+(2.32)$}\quad
&\quad\hspace{0.5mm}{$D_s^{*+}\gamma$} \quad 
& \quad\,  {$\sim 0.005$}($\ast$)\\
\hline  
\quad  {$\hat E^0(2.32)$}\quad 
&\quad\hspace{0.5mm} {$\langle{D\bar K}\rangle$} \quad 
& \quad( {weak int.})\quad\\
\hline
\end{tabular} \vspace{3mm}
\\
($\ast$) Discussed later.
\end{center}
%%%%%%%%%%%%%%%%%%%%%%%%%%%%%%%%%%%%%%%%%%%%%%%%%%%%%%%%%%%%%%%%%%%%%
%%%%%%%%%%%%%%%%%%%%%%%%%%%%%%%%%%%%%%%%%%%%%%%%%%%%%%%%%%%%%%%%%%%%%
%\vspace{2mm}
%\newpage

%%%%%%%%%%%%%%%%%%%%%%%%%%%%%%%%%
\section{Radiative Decays} 
%%%%%%%%%%%%%%%%%%%%%%%%%%%%%%%%%%

Now we study the radiative decays of $D_s^{*+}$ and $D_{s0}^+(2317)$ 
under the vector meson dominance (VMD). The results will be compared 
with $D_s^+\pi^0$ decays later. Amplitudes for the radiative decays of 
vector and scalar mesons are written as 
%%%%%%%%%%%%%%%%%%%%%%%%%%%%%%%%%%%%%%%%%%%%%%%%%%%%%%%%%%%%%%%%%%%%%
\begin{equation}
M(V\rightarrow P\gamma)=\epsilon^{\mu \nu \alpha \beta }
F_{\mu \nu }(\gamma)G_{\alpha \beta }(V) A(V\rightarrow P\gamma)
\label{eq:VPgamma}
\end{equation}
%%%%%%%%%%%%%%%%%%%%%%%%%%%%%%%%%%%%%%%%%%%%%%%%%%%%%%%%%%%%%%%%%%%%%
and 
%%%%%%%%%%%%%%%%%%%%%%%%%%%%%%%%%%%%%%%%%%%%%%%%%%%%%%%%%%%%%%%%%%%%%
\begin{equation}
M(S\rightarrow V\gamma)
=F_{\mu \nu }(\gamma)G^{\mu \nu }(V) A(S\rightarrow V\gamma), 
\label{eq:SVgamma}
\end{equation}
%%%%%%%%%%%%%%%%%%%%%%%%%%%%%%%%%%%%%%%%%%%%%%%%%%%%%%%%%%%%%%%%%%%%%
respectively. The amplitudes $A(V\rightarrow P\gamma)$ and 
$A(S\rightarrow V\gamma)$ are given by 
%%%%%%%%%%%%%%%%%%%%%%%%%%%%%%%%%%%%%%%%%%%%%%%%%%%%%%%%%%%%%%%%%%%%%
\begin{equation}
A(V\rightarrow P\gamma)=\sum_{V'=\rho,\,\omega,\,\phi,\,\psi}
{X_{V'}(0)\over m_{V'}^2}A(V\rightarrow PV') 
                                                   \label{eq:VpV'}
\end{equation}
%%%%%%%%%%%%%%%%%%%%%%%%%%%%%%%%%%%%%%%%%%%%%%%%%%%%%%%%%%%%%%%%%%%%%
and 
%%%%%%%%%%%%%%%%%%%%%%%%%%%%%%%%%%%%%%%%%%%%%%%%%%%%%%%%%%%%%%%%%%%%%
\begin{equation}
A(S\rightarrow V\gamma)
=\sum_{V'=\rho,\,\omega,\,\phi,\,\psi}
{X_{V'}(0)\over m_{V'}^2}A(S\rightarrow VV')
                                                   \label{eq:SVV'}
\end{equation}
%%%%%%%%%%%%%%%%%%%%%%%%%%%%%%%%%%%%%%%%%%%%%%%%%%%%%%%%%%%%%%%%%%%%%
under the VMD, where $X_V(0)$ is the $\gamma V$coupling strength on 
the photon-mass-shell. The OZI rule~\cite{OZI} selects possible vector 
mesons which can work as a pole in the radiative decays. 
The photon-momentum-square dependence of $X_V$ has been studied in 
Ref.~\cite{Terasaki-VMD}. The values of $X_V(0)$ have been measured by 
using photoproductions of vector mesons on various nuclei~\cite{Leith}. 
For the $\psi$ photoproduction, both the measured differential cross 
section at $t=0$, 
%%%%%%%%%%%%%%%%%%%%%%%%%%%%%%%%%%%%%%%%%%%%%%%%%%%%%%%%%%%%%%%%%%%%%%
$d\sigma(\gamma N\rightarrow \psi N)/dt|_{t=0}\simeq 20$ nb/GeV$^2$
around $\sim 20$ GeV, 
%%%%%%%%%%%%%%%%%%%%%%%%%%%%%%%%%%%%%%%%%%%%%%%%%%%%%%%%%%%%%%%%%%%%%%
and the $\psi N$ total cross section,  
%%%%%%%%%%%%%%%%%%%%%%%%%%%%%%%%%%%%%%%%%%%%%%%%%%%%%%%%%%%%%%%%%%%%%%
$\sigma_T(\psi N)=3.5\pm 0.8$ mb,  
%%%%%%%%%%%%%%%%%%%%%%%%%%%%%%%%%%%%%%%%%%%%%%%%%%%%%%%%%%%%%%%%%%%%%%%
measured in $A$-dependence of photoproduction cross 
sections~\cite{HLW} still have large uncertainties, where $t$, $N$ and 
$A$ are a momentum transfer squared, a nucleon and a mass number of 
nucleus. $\psi$ denotes the usual $J/\psi$. The values of $X_V(0)$ and 
$X_V(m_V^2)$ on the vector meson mass-shell are listed in Table~6, 
where $X_V(m_V^2)$ has been estimated from rates for leptonic decays 
of vector mesons. Their sign is determined by using the quark model. 
It should be noted that the above $|X_\phi(0)|$ and $|X_\psi(0)|$ are 
considerably smaller than $|X_\phi(m_\phi^2)|$ and $|X_\psi(m_\psi^2)|$, 
respectively.  
%%%%%%%%%%%%%%%%%%%%%%%%%%%%%%%%%%%%%%%%%%%%%%%%%%%%%%%%%%%%%%%%%%
\begin{center}
\begin{quote}
Table~6. Photon-vector meson coupling strengths in GeV$^2$. The sign of 
$X_V$ is determined by using the quark model. 
\end{quote}\vspace{3mm}

\begin{tabular}{c c c}
\hline
\quad $V$\quad &\quad 
{$X_V(k^2=0)$ in GeV$^2$}\quad 
&\quad $X_V(k^2=m_V^2)$ in GeV$^2$\quad
\\
\hline
\quad$\rho^0$\quad &\quad 
{$0.033\pm 0.003$}\quad 
&\quad $0.0357\pm 0.0008$\quad\\
\hline
\quad$\omega$\quad &\quad 
{$0.011\pm 0.001$}\quad 
&\quad $0.0109\pm 0.0002$\quad\\
\hline
\quad$\phi$\quad &\hspace{-0.5mm} 
{$-0.018\pm 0.004$}\quad 
&\hspace{1mm}$-0.0238\pm 0.0003$\quad\\
\hline
\quad $\psi$\quad &\quad 
{$\sim 0.054$}\quad 
& $0.380\pm 0.013$\quad\\
\hline
\end{tabular}
\end{center}
\vspace{3mm}
%%%%%%%%%%%%%%%%%%%%%%%%%%%%%%%%%%%%%%%%%%%%%%%%%%%%%%%%%%%%%%%%%%

Before going to the radiative decays of scalar mesons, we study 
radiative decays of vector mesons. We first estimate 
$A(\omega\rightarrow\pi^0\rho^0)$ as one of $A(V\rightarrow PV')$ 
coupling strengths from the measured rate for the 
$\omega\rightarrow\pi^0\gamma$~\cite{PDG04}, 
%%%%%%%%%%%%%%%%%%%%%%%%%%%%%%%%%%%%%%%%%%%%%%%%%%%%%%%%%%%%%%%%%%%%%
\begin{equation}
\Gamma(\omega\rightarrow\pi^0\gamma)_{\rm exp} = 0.734\pm 0.035 
                                                    \,\,{\rm MeV}. 
\end{equation}
%%%%%%%%%%%%%%%%%%%%%%%%%%%%%%%%%%%%%%%%%%%%%%%%%%%%%%%%%%%%%%%%%%%%%%
In this decay, the OZI-allowed vector meson pole is given by the 
$\rho^0$. 
Putting $V'=\rho^0$ in the amplitude, Eq.~(\ref{eq:VpV'}), inserting 
it to Eq.~(\ref{eq:VPgamma}) and using Eq.~(\ref{eq:rate-general}), 
we obtain 
%%%%%%%%%%%%%%%%%%%%%%%%%%%%%%%%%%%%%%%%%%%%%%%%%%%%%%%%%%%%%%%%%%%%%%%%
\begin{equation}
|A(\omega\rightarrow\pi^0\rho^0)|\simeq 18 \,\,({\rm GeV})^{-1}
\label{eq:omega-pi-rho}
\end{equation}
%%%%%%%%%%%%%%%%%%%%%%%%%%%%%%%%%%%%%%%%%%%%%%%%%%%%%%%%%%%%%%%%%%%%%%%
which will be used below as an input data. 

Now we study the $D_s^{*+}\rightarrow D_s^+\gamma$ decay comparing with 
the $\omega\rightarrow\pi^0\gamma$. The OZI-allowed poles are given by 
$\phi$ and $\psi$ mesons. Use of the $SU_f(4)$ relation of the 
$VPV'$ coupling strengths, 
%%%%%%%%%%%%%%%%%%%%%%%%%%%%%%%%%%%%%%%%%%%%%%%%%%%%%%%%%%%%%%%%%%%%%%
\begin{eqnarray}    
&&  
2A(D^{*0}\rightarrow D^0\rho^0)=2A(D^{*0}\rightarrow D^0\omega)
=\sqrt{2}A(D^{*0}\rightarrow D^0\psi)
\nonumber\\
&&  
=-2A(D^{*+}\rightarrow D^+\rho^0)=2A(D^{*+}\rightarrow D^+\omega)
=\sqrt{2}A(D^{*+}\rightarrow D^+\psi)
\nonumber\\
&&  
=\sqrt{2}A(D_s^{*+}\rightarrow D_s^+\phi)    
=\sqrt{2}A(D_s^{*+}\rightarrow D_s^+\psi)\,\,  
=A(\omega\rightarrow \pi^0\rho^0),
\label{eq:VPV'-SU_f(4)}   
\end{eqnarray}
%%%%%%%%%%%%%%%%%%%%%%%%%%%%%%%%%%%%%%%%%%%%%%%%%%%%%%%%%%%%%%%%%%%%%%%%%
leads to 
%%%%%%%%%%%%%%%%%%%%%%%%%%%%%%%%%%%%%%%%%%%%%%%%%%%%%%%%%%%%%%%%%%%%%%%%
\begin{equation}
\Gamma(D_s^{*+}\rightarrow D_s^+\gamma)_{SU_f(4)} 
                                      \simeq 0.8\,\,{\rm keV}
                                                 \label{eq:D_s^*-rad}, 
\end{equation}
%%%%%%%%%%%%%%%%%%%%%%%%%%%%%%%%%%%%%%%%%%%%%%%%%%%%%%%%%%%%%%%%%%%%%%%%
which is consistent with the experimental constraint~\cite{PDG04}, 
$\Gamma(D_s^{*+}\rightarrow D_s^+\gamma) < 1.8\,\,{\rm MeV}$. 

We here estimate uncertainties arising from the $SU_f(4)$ relation for 
the $VPV'$ coupling strengths. In the same way as the above, we can 
estimate rates for the $D^*\rightarrow D\gamma$ decays as 
%%%%%%%%%%%%%%%%%%%%%%%%%%%%%%%%%%%%%%%%%%%%%%%%%%%%%%%%%%%%%%%%%%%%%%%%
\begin{equation}
\Gamma(D^{*+}\rightarrow D^+\gamma)_{SU_f(4)}\simeq 2.4\,\,{\rm keV}
\quad {\rm and} \quad
\Gamma(D^{*0}\rightarrow D^0\gamma)_{SU_f(4)}\simeq 19\,\,{\rm keV},
\end{equation}
%%%%%%%%%%%%%%%%%%%%%%%%%%%%%%%%%%%%%%%%%%%%%%%%%%%%%%%%%%%%%%%%%%%%%%%%
which can be compared with the measured ones~\cite{PDG04}, 
%%%%%%%%%%%%%%%%%%%%%%%%%%%%%%%%%%%%%%%%%%%%%%%%%%%%%%%%%%%%%%%%%%%%%%%%
\begin{equation}
\Gamma(D^{*+}\rightarrow D^+\gamma)_{\rm exp} = 1.5\pm 0.5\,\,
{\rm keV}\quad {\rm and}\quad
\Gamma(D^{*0}\rightarrow D^0\gamma)_{\rm exp} < 800\,\,{\rm keV}. 
\end{equation}
%%%%%%%%%%%%%%%%%%%%%%%%%%%%%%%%%%%%%%%%%%%%%%%%%%%%%%%%%%%%%%%%%%%%%%%%
In particular, the ratio 
%%%%%%%%%%%%%%%%%%%%%%%%%%%%%%%%%%%%%%%%%%%%%%%%%%%%%%%%%%%%%%%%%%%%%%%%
\begin{equation}
{\Gamma(D^{*+}\rightarrow D^+\gamma)_{SU_f(4)} \over 
\Gamma(D^{*+}\rightarrow D^+\gamma)_{\rm exp}}  \simeq 1.6  
\end{equation}
%%%%%%%%%%%%%%%%%%%%%%%%%%%%%%%%%%%%%%%%%%%%%%%%%%%%%%%%%%%%%%%%%%%%%%%%
implies that the $SU_f(4)$ symmetry again overestimates the rate for 
radiative decays by $\sim 60\,\%$ with large errors. 

The radiative $D_{s0}^+(2317)\rightarrow D_s^{*+}\gamma$ decay is now in 
order. Possible three cases, i.e., 
(i) $D_{s0}^+(2317)$ is assumed to be the iso-singlet four-quark meson 
$\hat F_0^+$, 
(ii) $D_{s0}^+(2317)$ is assumed to be the conventional scalar meson 
$D_{s0}^{*+}$, and 
(iii) $D_{s0}^+(2317)$ is assumed to be the iso-triplet four-quark meson 
$\hat F_I^+$, are studied.  
In the case (iii), the result will be compared with the rate, 
$\Gamma(\hat F_I^+\rightarrow D_s^+\pi^0)_{SU_f(4)}\sim 9$ MeV 
in Eq.~(\ref{eq:rate-of-F_I-SU(4)}) or 
$\Gamma(\hat F_I^+\rightarrow D_s^+\pi^0)\sim 4 - 5$ MeV, in 
Eq.~(\ref{eq:rate-of-F_I}) and it will be seen that the experiments favor 
the assignment of $D_{s0}^+(2317)$ to $\hat F_I^+$. In the cases (i) 
and (ii), the results will be compared with the isospin non-conserving 
decays in the next section. We, first, consider the case 
(iii) $D_{s0}^+(2317)=\hat F_I^+\sim [cn][\bar s\bar n]_{I=1}$. 
In the $\hat F_I^+\rightarrow D_s^{*+}\gamma$, the OZI-allowed pole is 
given by $\rho^0$. The $SU_f(4)$ relation for the $SVV'$ couplings is 
%%%%%%%%%%%%%%%%%%%%%%%%%%%%%%%%%%%%%%%%%%%%%%%%%%%%%%%%%%%%%%%%%
\begin{equation}
2A(\hat F_I^+\rightarrow D_s^{*+}\rho^0) 
= 2A(\hat F_0^+\rightarrow D_s^{*+}\omega) 
= A(\phi\rightarrow \hat\delta^{s0}\rho^0)(\beta')_{SU_f(4)}, 
                                               \label{eq:SVV'-SU(4)}
\end{equation}
%%%%%%%%%%%%%%%%%%%%%%%%%%%%%%%%%%%%%%%%%%%%%%%%%%%%%%%%%%%%%%%%%%%%%%%%
where the overlapping factor is now $\beta'$ (in place of $\beta$) 
because the mesons which couple to the $[cq][\bar q\bar q]$ meson are 
now two vector mesons. Use of the measured rate~\cite{PDG04}, 
%%%%%%%%%%%%%%%%%%%%%%%%%%%%%%%%%%%%%%%%%%%%%%%%%%%%%%%%%%%%%%%%%%%%%%%%
\begin{equation}
\Gamma(\phi\rightarrow a_0(980)\gamma)_{\rm exp} = 0.32\pm 0.03
\,\,{\rm keV},                                   \label{eq:SVV'-input}
\end{equation}
%%%%%%%%%%%%%%%%%%%%%%%%%%%%%%%%%%%%%%%%%%%%%%%%%%%%%%%%%%%%%%%%%%%%%%%%
as the input data and the $SU_f(4)$ relation, Eq.~(\ref{eq:SVV'-SU(4)}), 
with $|\beta'|_{SU_f(4)}^2\simeq 1/4$ provides 
%%%%%%%%%%%%%%%%%%%%%%%%%%%%%%%%%%%%%%%%%%%%%%%%%%%%%%%%%%%%%%%%%%%%%%%% 
\begin{equation}
\Gamma(\hat F_I^+\rightarrow D_s^{*+}\gamma)_{SU_f(4)} \simeq 45 
\,\,{\rm keV}.                                  \label{eq:F_I-rad}
\end{equation}
%%%%%%%%%%%%%%%%%%%%%%%%%%%%%%%%%%%%%%%%%%%%%%%%%%%%%%%%%%%%%%%%%%%%%%%% 
Then its ratio to the rate  
%%%%%%%%%%%%%%%%%%%%%%%%%%%%%%%%%%%%%%%%%%%%%%%%%%%%%%%%%%%%%%%%%%%%%%%% 
$\Gamma(\hat F_I^+\rightarrow D_s^+\pi^0)_{SU_f(4)}\sim 9$ MeV 
%%%%%%%%%%%%%%%%%%%%%%%%%%%%%%%%%%%%%%%%%%%%%%%%%%%%%%%%%%%%%%%%%%%%%%%% 
in Eq.~(\ref{eq:rate-of-F_I-SU(4)}) %{eq:rate-of-F_I}
%%%%%%%%%%%%%%%%%%%%%%%%%%%%%%%%%%%%%%%%%%%%%%%%%%%%%%%%%%%%%%%%%%%%%%%% 
is 
%%%%%%%%%%%%%%%%%%%%%%%%%%%%%%%%%%%%%%%%%%%%%%%%%%%%%%%%%%%%%%%%%%%%%%%% 
$R(\hat F_I^+)_{SU_f(4)} \sim 0.005$.         
%%%%%%%%%%%%%%%%%%%%%%%%%%%%%%%%%%%%%%%%%%%%%%%%%%%%%%%%%%%%%%%%%%%%%%%%
It will be left intact even if the $SU_f(4)$ symmetry breaking arising 
from overlap of spatial wavefunctions which has been discussed to 
be approximately the same in both cases is taken into account, i.e., 
%%%%%%%%%%%%%%%%%%%%%%%%%%%%%%%%%%%%%%%%%%%%%%%%%%%%%%%%%%%%%%%%%%%%%%%% 
\begin{equation}
R(\hat F_I^+) =
{\Gamma(\hat F_I^+\rightarrow D_s^{*+}\gamma)
\over \Gamma(\hat F_I^+\rightarrow D_s^+\pi^0)}\sim 0.005,         
                                                 \label{eq:F_I-ratio}
\end{equation}
%%%%%%%%%%%%%%%%%%%%%%%%%%%%%%%%%%%%%%%%%%%%%%%%%%%%%%%%%%%%%%%%%%%%%%%% 
which satisfies well the experimental constraint, Eq.~(\ref{eq:CLEO}).  
Therefore the experiment favors the assignment of the new resonance 
$D_{s0}^+(2317)$ to the $I_3=0$ component $\hat F_I^+$ of the 
iso-triplet four-quark mesons. 

Next, we study radiative decays of iso-singlet scalar mesons in the 
cases (i) and (ii). %, and their results will be compared with isospin 
%non-conserving $D_s^+\pi^0$ decays in the next section. 
In the case 
(i) $D_{s0}^+(2317)=\hat F_0^+\sim [cn][\bar s\bar n]_{I=0}$, 
the OZI-allowed pole is given by $\omega$. When the $SU_f(4)$ relation, 
Eq.~(\ref{eq:SVV'-SU(4)}), with $|\beta'|_{SU_f(4)}^2\simeq 1/4$ and 
Eq.~(\ref{eq:SVV'-input}) as the input data are taken, the resulting 
rate for the radiative decay of $\hat F_0^+$ is 
%%%%%%%%%%%%%%%%%%%%%%%%%%%%%%%%%%%%%%%%%%%%%%%%%%%%%%%%%%%%%%%%%%%%%%%% 
\begin{equation}
\Gamma(\hat F_0^+\rightarrow D_s^{*+}\gamma)_{SU_f(4)} \simeq 4.7 
\,\,{\rm keV},                                  \label{eq:rate-rad-F_0}
\end{equation}
%%%%%%%%%%%%%%%%%%%%%%%%%%%%%%%%%%%%%%%%%%%%%%%%%%%%%%%%%%%%%%%%%%%%%%%%
which is smaller by about an order than the above 
$\Gamma(\hat F_I^+\rightarrow D_s^{*+}\gamma)_{SU_f(4)}\simeq 45$ keV 
because of $X_{\rho^0}(0) \simeq 3X_{\omega}(0)$. 
In the case (ii) $D_{s0}^+(2317) = D_{s0}^{*+}\sim \{c\bar s\}$, 
the OZI-allowed poles are given by $\phi$ and $\psi$ mesons.  
The $SU_f(4)$ relation of the $SVV'$ couplings is now 
%%%%%%%%%%%%%%%%%%%%%%%%%%%%%%%%%%%%%%%%%%%%%%%%%%%%%%%%%%%%%%%%%%%%%%%%
\begin{equation}
2A(D_{s0}^{*+}\rightarrow D_s^{*+}\phi)
=2A(D_{s0}^{*+}\rightarrow D_s^{*+}\psi)
=A(\chi_{c0}\rightarrow \psi\psi),               \label{eq:SVV'-conv}
\end{equation}
%%%%%%%%%%%%%%%%%%%%%%%%%%%%%%%%%%%%%%%%%%%%%%%%%%%%%%%%%%%%%%%%%%%%%%%%
where $\chi_{c0}$ denotes the $^3P_0\,\,\{c\bar c\}$. 
The $SU_f(4)$ relation, Eq.~(\ref{eq:SVV'-conv}), reads  
%%%%%%%%%%%%%%%%%%%%%%%%%%%%%%%%%%%%%%%%%%%%%%%%%%%%%%%%%%%%%%%%%%%%%%%%
\begin{equation}
\Gamma(D_{s0}^{*+}\rightarrow D_s^{*+}\gamma)_{SU_f(4)}
\sim 35\,\,{\rm keV}.  
                                                 \label{eq:rate-rad-conv}
\end{equation}
%%%%%%%%%%%%%%%%%%%%%%%%%%%%%%%%%%%%%%%%%%%%%%%%%%%%%%%%%%%%%%%%%%%%%%%%
Here 
%%%%%%%%%%%%%%%%%%%%%%%%%%%%%%%%%%%%%%%%%%%%%%%%%%%%%%%%%%%%%%%%%%%%%%%%
$\Gamma(\chi_{c0}\rightarrow \psi\gamma)_{\rm exp}=119\pm 15$ 
keV~\cite{PDG04} 
%%%%%%%%%%%%%%%%%%%%%%%%%%%%%%%%%%%%%%%%%%%%%%%%%%%%%%%%%%%%%%%%%%%%%%%%
has been taken as the input data. The above result is much larger than 
the one of the constituent quark models~\cite{BEH,AG} in which the rate 
for the radiative decay has been strongly suppressed by taking a large 
strange ($s$) quark mass, $m_c\simeq (3-4)m_s$, far from the heavy charm 
quark picture ($m_c\gg m_s$). Such a "fat" $s$-quark reduces drastically 
the dipole moment of the $\{c\bar s\}$ system. However, it might be 
unnatural since a more slim $s$-quark, $m_s\simeq 0.9$ GeV, seems to be 
favored by a semi-relativistic quark model~\cite{MM} in the 
$\{c\bar s\}$ 
meson spectroscopy and by the recent result (at $\sim 2$ GeV scale) 
from the lattice QCD~\cite{Gupta} as mentioned before. If $m_c\gg m_s$, 
the rate for the radiative decay will be much larger than the ones in 
Refs.~\cite{BEH} and \cite{AG}, although, in this case, the $s$-quark 
might not behave non-relativistically. We will compare the results in 
Eqs.~(\ref{eq:rate-rad-F_0}) and (\ref{eq:rate-rad-conv}) 
with rates for the isospin non-conserving decays in the next section. 

%%%%%%%%%%%%%%%%%%%%%%%%%%%%%%%%%%%%%%%%%%%%
\section{Isospin non-conserving decays}
%%%%%%%%%%%%%%%%%%%%%%%%%%%%%%%%%%%%%%%%%%%%

When $D_{s0}^+(2317)$ is assigned to an iso-singlet state, 
the $D_{s0}^+(2317) \rightarrow D_s^+\pi^0$ decay does not conserve 
isospin. It is assumed to proceed through a tiny $\pi^0$-$\eta$ 
mixing as usual~\cite{Dalitz,CW}, where the physical $(\pi^0)_{\rm phys}$ 
is an admixture of the pure iso-triplet $\pi^0$ and the iso-singlet 
$\eta$, i.e., 
%%%%%%%%%%%%%%%%%%%%%%%%%%%%%%%%%%%%%%%%%%%%%%%%%%%%%%%%%%%%%%%%%%%%%%%
\begin{equation}
\pi^0_{\rm phys}\simeq \pi^0 + \epsilon \eta, \quad (|\epsilon|\ll 1).  
\end{equation}
%%%%%%%%%%%%%%%%%%%%%%%%%%%%%%%%%%%%%%%%%%%%%%%%%%%%%%%%%%%%%%%%%%%%%%%
To use the hard pion technique in the IMF, we replace the above mixing 
by 
%%%%%%%%%%%%%%%%%%%%%%%%%%%%%%%%%%%%%%%%%%%%%%%%%%%%%%%%%%%%%%%%%%%%%%%%
\begin{equation}
(A_{\pi^0})_{\rm phys} \simeq A_{\pi^0} + \epsilon A_{\eta}, 
                         \quad (|\epsilon|\ll 1).     \label{eq:eta-pi}
\end{equation}
%%%%%%%%%%%%%%%%%%%%%%%%%%%%%%%%%%%%%%%%%%%%%%%%%%%%%%%%%%%%%%%%%%%%%%%%
The isospin non-conserving parameter $\epsilon$ is estimated below. 
The non-diagonal element of $\pi^0$-$\eta$ mass matrix was given by
%%%%%%%%%%%%%%%%%%%%%%%%%%%%%%%%%%%%%%%%%%%%%%%%%%%%%%%%%%%%%%%%%%%%%%%
\begin{equation}
\langle{\pi^0|\delta m^2|\eta}\rangle 
= \sqrt{1\over 3}\Bigl\{(m_{\pi^0}^2 - m_{\pi^+}^2) 
                        - (m_{K^0}^2 - m_{K^+}^2)\Bigr\}
\end{equation}
%%%%%%%%%%%%%%%%%%%%%%%%%%%%%%%%%%%%%%%%%%%%%%%%%%%%%%%%%%%%%%%%%%%%%%%%
long time ago~\cite{Okubo-Sakita} by assuming that the mixing is caused 
by the electromagnetic interactions.  Then the $\pi^0$-$\eta$ mixing 
parameter was estimated as~\cite{Dalitz}  
%%%%%%%%%%%%%%%%%%%%%%%%%%%%%%%%%%%%%%%%%%%%%%%%%%%%%%%%%%%%%%%%%%%%%%%
\begin{equation}
\epsilon = -{\langle{\pi^0|\delta m^2|\eta}\rangle 
\over m_\eta^2 - m_{\pi^0}^2} = 0.0105 \pm 0.0013,  
                                             \label{eq:epsilon}
\end{equation}
%%%%%%%%%%%%%%%%%%%%%%%%%%%%%%%%%%%%%%%%%%%%%%%%%%%%%%%%%%%%%%%%%%%%%%%%
by inserting the well-known mass values of the related mesons. The 
estimated value of $\epsilon$ is of the order of the fine structure 
constant $\alpha$, i.e., $\epsilon \sim O(\alpha)$ as expected. 
It implies that the isospin non-conserving interactions are much weaker 
than the electromagnetic ones. 

With the above value of $\epsilon$, we estimate the rate for the 
isospin non-conserving $D_s^{*+}\rightarrow D_s^+\pi^0$ decay. 
The $\pi^0$-$\eta$ mixing, the OZI rule and the asymptotic $SU_f(4)$ 
symmetry lead to 
%%%%%%%%%%%%%%%%%%%%%%%%%%%%%%%%%%%%%%%%%%%%%%%%%%%%%%%%%%%%%%%%%%%%%%%
\begin{equation}
\langle{D_s^+|(A_{\pi^0})_{\rm phys}|D_s^{*+}}\rangle
\simeq -{1\over 2}\epsilon\sin\Theta 
                      \langle{\pi^+|A_{\pi^+}|\rho^0}\rangle,
\end{equation}
%%%%%%%%%%%%%%%%%%%%%%%%%%%%%%%%%%%%%%%%%%%%%%%%%%%%%%%%%%%%%%%%%%%%
where $\eta=\cos\Theta \eta^n - \sin\Theta \eta^s$ because of the 
$\eta$-$\eta'$ mixing. Using 
$|\langle{\pi^+|A_{\pi^+}|\rho^0\rangle}|\simeq 1.0$ 
from 
$\Gamma(\rho\rightarrow\pi\pi)_{\rm exp}\simeq 150$ MeV~\cite{PDG04}, 
$\Theta\simeq 35^\circ$ as before, and $\epsilon\simeq 0.0105$ 
estimated above, we obtain 
%%%%%%%%%%%%%%%%%%%%%%%%%%%%%%%%%%%%%%%%%%%%%%%%%%%%%%%%%%%%%%%%%%%
\begin{equation}
\Gamma(D_s^{*+}\rightarrow D_s^+\pi^0)_{SU_f(4)}\simeq 50\,\,{\rm eV} 
                                         \label{eq:isospin-noncons}
\end{equation}
%%%%%%%%%%%%%%%%%%%%%%%%%%%%%%%%%%%%%%%%%%%%%%%%%%%%%%%%%%%%%%%%%%%
which is much smaller than the 
%%%%%%%%%%%%%%%%%%%%%%%%%%%%%%%%%%%%%%%%%%%%%%%%%%%%%%%%%%%%%%%%%%%
$\Gamma(D_s^{*+}\rightarrow D_s^+\gamma)_{SU_f(4)}\simeq 0.8$ keV 
%%%%%%%%%%%%%%%%%%%%%%%%%%%%%%%%%%%%%%%%%%%%%%%%%%%%%%%%%%%%%%%%%%%
estimated in Eq.~(\ref{eq:D_s^*-rad}) before. 
The ratio of these two rates is given by 
%%%%%%%%%%%%%%%%%%%%%%%%%%%%%%%%%%%%%%%%%%%%%%%%%%%%%%%%%%%%%%%%%%%%%
\begin{equation}
R(D_s^{*+})^{-1}\simeq 0.06
\end{equation}
%%%%%%%%%%%%%%%%%%%%%%%%%%%%%%%%%%%%%%%%%%%%%%%%%%%%%%%%%%%%%%%%%%%%%%%
which satisfy well the measured ratio 
in Eq.~(\ref{eq:constraints}).

Next, we consider isospin non-conserving decays of scalar mesons. 
We study possible two cases, (i) $D_{s0}^+(2317)$ as the iso-singlet 
four-quark meson $\hat F_0^+$, and (ii) $D_{s0}^+(2317)$ as the 
conventional $D_{s0}^{*+}$ meson. 
In the case (i), the $\pi^0$-$\eta$ mixing and the OZI rule lead to 
%%%%%%%%%%%%%%%%%%%%%%%%%%%%%%%%%%%%%%%%%%%%%%%%%%%%%%%%%%%%%%%%%%%%%%%%
\begin{equation}
\langle{D_s^+|(A_{\pi^0})_{\rm phys}|\hat F_0^+}\rangle
\simeq 
\epsilon \cos\Theta \langle{D_s^+|A_{\eta^n}|\hat F_0^+}\rangle,
\end{equation}
%%%%%%%%%%%%%%%%%%%%%%%%%%%%%%%%%%%%%%%%%%%%%%%%%%%%%%%%%%%%%%%%%%%%%%%
where $\Theta\simeq 35^\circ$ and $\epsilon\simeq 0.0105$  as before.  
The asymptotic $SU_f(4)$ relates 
$\langle{D_s^+|A_{\eta^n}|\hat F_0^+}\rangle$ to 
$\langle{\eta^s|A_{\pi^-}|\hat\delta^{s+}}\rangle$ 
whose size has been estimated in Eq.~(\ref{eq:4-quark-coupling}), i.e., 
%%%%%%%%%%%%%%%%%%%%%%%%%%%%%%%%%%%%%%%%%%%%%%%%%%%%%%%%%%%%%%%%%%%%%%%% 
\begin{equation}
2\langle{D_s^+|A_{\eta^n}|\hat F_0^+}\rangle
=\sqrt{1\over 2}\langle{\eta^s|A_{\pi^-}|\hat\delta^{s+}}\rangle\beta. 
                                                \label{eq:I-viol-SU(4)}
\end{equation}
%%%%%%%%%%%%%%%%%%%%%%%%%%%%%%%%%%%%%%%%%%%%%%%%%%%%%%%%%%%%%%%%%%%%%%%
Using Eq.~(\ref{eq:I-viol-SU(4)}) with Eq.~(\ref{eq:4-quark-coupling}) 
and $|\beta|_{SU_f(4)}^2\simeq 1/12$, we get 
%%%%%%%%%%%%%%%%%%%%%%%%%%%%%%%%%%%%%%%%%%%%%%%%%%%%%%%%%%%%%%%%%%%%%%%% 
\begin{equation}
\Gamma(\hat F_0^+\rightarrow D_s^+\pi^0)_{SU_f(4)}\sim 0.7\,\,{\rm keV}, 
\end{equation}
%%%%%%%%%%%%%%%%%%%%%%%%%%%%%%%%%%%%%%%%%%%%%%%%%%%%%%%%%%%%%%%%%%%%%%%
which is much smaller than the rate for the radiative decay in 
Eq.~(\ref{eq:rate-rad-F_0}) as expected. In the case (ii), where 
$D_{s0}^+(2317)$ is assigned to the conventional scalar meson 
$D_{s0}^{*+}$, the $\pi^0$-$\eta$ mixing and the OZI rule lead to 
%%%%%%%%%%%%%%%%%%%%%%%%%%%%%%%%%%%%%%%%%%%%%%%%%%%%%%%%%%%%%%%%%%%%%%%%
\begin{equation}
\langle{D_s^+|(A_{\pi^0})_{\rm phys}|D_{s0}^{*+}}\rangle
\simeq 
-\epsilon \sin\Theta \langle{D_s^+|A_{\eta^s}|D_{s0}^{*+}}\rangle,
                                              \label{eq:eta-pi-conv}
\end{equation}
%%%%%%%%%%%%%%%%%%%%%%%%%%%%%%%%%%%%%%%%%%%%%%%%%%%%%%%%%%%%%%%%%%%%%%%
where $\Theta\simeq 35^\circ$ and $\epsilon\simeq 0.0105$ again. 
The asymptotic $SU_f(4)$ symmetry relates the asymptotic 
matrix element $\langle{D_s^+|A_{\eta^s}|D_{s0}^{*+}}\rangle$ to 
the experimentally known $\langle{K^+|A_{\pi^+}|K_0^{*0}(1430)}\rangle$, 
%%%%%%%%%%%%%%%%%%%%%%%%%%%%%%%%%%%%%%%%%%%%%%%%%%%%%%%%%%%%%%%%%%%%%%%% 
\begin{equation}
\langle{D_s^+|A_{\eta^s}|D_{s0}^{*+}}\rangle
=\langle{K^+|A_{\pi^+}|K_0^{*0}(1430)}\rangle, 
                                              \label{eq:SU(4)-conv}
\end{equation}
%%%%%%%%%%%%%%%%%%%%%%%%%%%%%%%%%%%%%%%%%%%%%%%%%%%%%%%%%%%%%%%%%%%%%%%
where $K_0^{*+}(1430)$ has been assigned to the conventional scalar 
$\{n\bar s\}$ meson as usual~\cite{CT}. The size of the matrix element 
in the r.h.s. of the above equation is estimated as 
%%%%%%%%%%%%%%%%%%%%%%%%%%%%%%%%%%%%%%%%%%%%%%%%%%%%%%%%%%%%%%%%%%%%%%%% 
\begin{equation}
|\langle{K^+|A_{\pi^+}|K_0^{*0}(1430)}\rangle|\simeq 0.29
                                                  \label{eq:AME-conv}
\end{equation}
%%%%%%%%%%%%%%%%%%%%%%%%%%%%%%%%%%%%%%%%%%%%%%%%%%%%%%%%%%%%%%%%%%%%%%%
by using the mass $m_{K_0^*}=1412\pm 6$ MeV and the measured rate 
$\Gamma(K_0^{*0}(1430)\rightarrow K^+\pi^-)_{\rm exp}=182\pm 24$ MeV  
from 
%%%%%%%%%%%%%%%%%%%%%%%%%%%%%%%%%%%%%%%%%%%%%%%%%%%%%%%%%%%%%%%%%%%%
$\Gamma_{K_0^*}=294\pm 23$ MeV and 
$Br(K_0^{*0}\rightarrow K~+\pi^-) = 93\pm 10$ \% 
%%%%%%%%%%%%%%%%%%%%%%%%%%%%%%%%%%%%%%%%%%%%%%%%%%%%%%%%%%%%%%%%%%%%
compiled by the PDG04~\cite{PDG04}.  
With help of Eqs.~(\ref{eq:SU(4)-conv}) and (\ref{eq:AME-conv}), 
the rate for the $D_{s0}^{*+}\rightarrow D_s^+\pi^0$ is estimated as
%%%%%%%%%%%%%%%%%%%%%%%%%%%%%%%%%%%%%%%%%%%%%%%%%%%%%%%%%%%%%%%%%%%%%%%%
\begin{equation}
\Gamma(D_{s0}^{*+}\rightarrow D_s^+\pi^0)_{SU_f(4)}
                                         \simeq 0.6 \,\,{\rm keV} ,
                                              \label{eq:rate-conv}
\end{equation}
%%%%%%%%%%%%%%%%%%%%%%%%%%%%%%%%%%%%%%%%%%%%%%%%%%%%%%%%%%%%%%%%%%%%%%%
which is again much smaller than the rate for the radiative decay of 
$D_{s0}^{*+}$, Eq.~(\ref{eq:rate-rad-conv}), as expected. 

Here we summarize the ratios of rates for the radiative $D_s^{*+}\gamma$ 
decay to the $D_s^+\pi^0$ in Table 7. The experimental data to be 
compared is $R(D_{s0}^+)_{\rm exp} < 0.059$ in Eq.~(\ref{eq:CLEO}). 
As seen in Table~7, $R(\hat F_0^+)$ and $R(D_{s0}^{*+})$ are much 
larger than the experimental upper bound. However, it is quite natural 
as discussed before. Therefore, the assignment of the iso-singlet state 
($\hat F_0^+$ or $D_{s0}^{*+}$) will be excluded, although there have 
been many efforts~\cite{BEH,Colangello,MS,Zhu} to reconcile the 
assignments with the experimental constraint, Eq.(\ref{eq:CLEO}). 
In fact, it is very strange that the ratio is predicted to be much 
smaller than unity when $D_{s0}^+(2317)$ is assigned to the conventional 
$D_{s0}^{*+}$, i.e., it means that a higher order term overcomes a lower 
order term in perturbation theory, since the rate for the isospin 
non-conserving decay is proportional to $|\epsilon|^2\sim O(\alpha^2)$ 
while the one for the radiative decay is of the order of $\alpha$. 
When $D_{s0}^+(2317)$ is assigned to the iso-triplet 
$\hat F_I^+\sim [cn][\bar s\bar n]_{I=1}$, however, the ratio 
$R(\hat F_I^+)$ satisfies well the experimental constraint. 
It is again quite natural since the electromagnetic interactions 
are much weaker than the isospin conserving strong interactions. 
Therefore, the assignment of $D_{s0}^+(2317)$ to $\hat F_I^+$ 
is favored by the experiment, while the assignments to the 
iso-singlet states are inconsistent with it. 
%%%%%%%%%%%%%%%%%%%%%%%%%%%%%%%%%%%%%%%%%%%%%%%%%%%%%%%%%%%%%%%%
%\newpage
%%%%%%%%%%%%%%%%%%
\begin{center}
\begin{quote}
Table~7. Ratio of the rates for the radiative $D_s^{*+}\gamma$ decay to 
the $D_s^+\pi^0$ of the scalar ($S=\hat F_I^+,\,\hat F_0^+$ and 
$D_{0s}^{*+}$) mesons. The experimental data 
is given in Ref.~\cite{CLEO-D_s}.  
\end{quote}
\vspace{2mm}
\begin{tabular}
{c c c c}
\hline
Scalar meson (S) & $\hat F_I^+$ & $\hat F_0^+$ &$D_{s0}^{*+}$ \\
\hline
$R(S)$ & $\sim 0.005$ & $\sim 7$ & $\sim 50$\\
\hline
$R(D_{s0}^+)_{\rm exp}$ &  & $< 0.059$ &  \\
\hline
\end{tabular}
\end{center}
\vspace{2mm}
%%%%%%%%%%%%%%%%%%%%%%%%%%%%%%%%%%%%%%%%%%%%%%%%%%%%%%%%%%%%%%%%%%%

%%%%%%%%%%%%%%%%%%%%%%%%%%%%%%%%%%%%%%%%%%%%%%%%%%
\section{Conventional scalar mesons with charm}
%%%%%%%%%%%%%%%%%%%%%%%%%%%%%%%%%%%%%%%%%%%%%%%%%%

We have studied scalar four-quark $[cq][\bar s\bar q]$ mesons with charm 
in the previous sections. In this section we study the conventional 
charm scalar mesons, $D_{0}^*\sim \{c\bar n\}$ and 
$D_{s0}^{*+}\sim \{c\bar s\}$, comparing with the light strange scalar 
meson $K_0^*$ which is usually assigned to the scalar $\{n\bar s\}$ 
meson~\cite{CT}. 

Since the mass of $D_0^*$ has not definitely been known, we take 
tentatively $m_{D_0^*}\simeq 2.35$ GeV which is close to the average 
of the central values of the broad bumps in the $D\pi$ mass 
distributions and is consistent with predictions from various 
approaches, for example, potential models~\cite{GK,potential},  
lattice QCD simulations~\cite{quench,Bali,UKQCD}, calculations 
based on QCD sum rule~\cite{HT-QCDSR}, etc. 

The asymptotic $SU_f(4)$ symmetry relates asymptotic matrix elements 
of axial charges, $A_\pi$ and $A_K$, taken between a pseudoscalar 
meson state and a conventional scalar state to each other, for example, 
%%%%%%%%%%%%%%%%%%%%%%%%%%%%%%%%%%%%%%%%%%%%%%%%%%%%%%%%%%%%%%%%%%%%%%%
\begin{equation}
\langle{D^+|A_{\pi^+}|D_0^{*0}}\rangle 
=\langle{D^0|A_{K^-}|D_{s0}^{*+}}\rangle
=\langle{K^+|A_{\pi^+}|K_{0}^{*0}}\rangle, 
                                                 \label{eq:AME-SU(4)}
\end{equation}
%%%%%%%%%%%%%%%%%%%%%%%%%%%%%%%%%%%%%%%%%%%%%%%%%%%%%%%%%%%%%%%%%%%%%%%
which are used below.  The size of the last matrix element has 
already been estimated in Eq.~(\ref{eq:AME-conv}). 
Use of the $SU_f(4)$ relation, Eq.~(\ref{eq:AME-SU(4)}), with the isospin 
$SU_I(2)$ symmetry and Eq.~(\ref{eq:rate-general}) with 
Eq.~(\ref{eq:hard-pion}) leads us to the following width of $D_0^*$ which 
is dominated by the $D_0^*\rightarrow D\pi$ decays,  
%%%%%%%%%%%%%%%%%%%%%%%%%%%%%%%%%%%%%%%%%%%%%%%%%%%%%%%%%%%%%%%%%%%%%%%,  
$(\Gamma_{D_0^*})_{SU_f(4)}
\simeq \Gamma(D_0^*\rightarrow D\pi)_{SU_f(4)} \simeq 90$ MeV. 
%%%%%%%%%%%%%%%%%%%%%%%%%%%%%%%%%%%%%%%%%%%%%%%%%%%%%%%%%%%%%%%%%%%%%%%
When the $SU_f(4)$ symmetry breaking is taken into account, 
the results will be reduced by about $50$ \%, i.e., 
%%%%%%%%%%%%%%%%%%%%%%%%%%%%%%%%%%%%%%%%%%%%%%%%%%%%%%%%%%%%%%%%%%%%%%%
\begin{equation} 
 \Gamma_{D_0^*}\sim 50 \,\,{\rm MeV}.       \label{eq:width-D_0}
\end{equation}
%%%%%%%%%%%%%%%%%%%%%%%%%%%%%%%%%%%%%%%%%%%%%%%%%%%%%%%%%%%%%%%%%%%%%%%
The above width $\Gamma_{D_0^*}$ is still rather broad but much narrower 
than the widths of the bumps in the $D\pi$ mass distributions observed by 
the BELLE and by the FOCUS collaboration. Therefore, we 
propose~\cite{Terasaki-mquark,Terasaki-D_0,TM} that they should be 
re-interpreted as four-quark mesons and conventional mesons. 
We do so because of the evidence that the charm-strange scalar meson 
$D_{s0}^+(2317)$ should be regarded as a four-quark state (as studied 
previous sections) and because there would be no room of charm 
non-strange scalar four-quark mesons if each bump were saturated by 
the conventional $D_0^*$. In addition, we expect $m_{D_0^*}$ 
to be in this mass region below 2460 MeV~\cite{HT-QCDSR} in 
consistency with other approaches as discussed before. 
Therefore, we consider that the bump has a structure comprising 
a broader ($\Gamma_{D_0^*}\sim 50 -90$ MeV) conventional $D_0^{*}$ 
in the region of its upper half and a narrower 
($\Gamma_{\hat D} \sim 5 - 10$ MeV) four-quark $\hat D$ in the region 
of its lower tail. 

For charm-strange scalar mesons, we have discussed $D_{s0}^+(2317)$ 
as a possible evidence for existence of charmed four-quark mesons in 
the previous sections. Its width is very narrow; the measured one has 
been compiled as $\Gamma_{D_{s0}^+(2317)} < 4.7$ MeV and our estimate 
has been  $\Gamma_{\hat F_I} \sim 4 - 5$ MeV. 
When $m_{D_{0}^*}\simeq 2.35$ GeV is taken, the mass of strange 
counterpart, $D_{s0}^*$, of the conventional $D_{0}^*$ is crudely 
estimated as $m_{D_{s0}^*}\simeq 2.45$ GeV by using a simple quark 
counting. Here we take the above value of $D_{s0}^*$, although there 
have been efforts~\cite{MM,Narison,YSKK} to reconcile its value with 
the measured $m_{D_{s0}}\simeq 2317$ MeV. 
This is because assigning $D_{s0}^+(2317)$ to the scalar $\{c\bar s\}$ 
meson is inconsistent with the experimental constraints, 
Eq.~(\ref{eq:constraints}), as seen in the previous section. 
Our $m_{D_{s0}^*}\simeq 2.45$ GeV is beyond the threshold of the $DK$ 
decays. 
Using a hard kaon approximation ($m_K^2\rightarrow 0$) 
in place of the hard pion ($m_\pi^2\rightarrow 0$), we can estimate 
the width of $D_{s0}^*$ which is dominated by the 
$D_{s0}^{*+}\rightarrow (DK)^+$ decays, i.e., 
%%%%%%%%%%%%%%%%%%%%%%%%%%%%%%%%%%%%%%%%%%%%%%%%%%%%%%%%%%%%%%%%%%%%%%%
$(\Gamma_{D_{s0}^*})_{SU_f(4)}
\simeq \Gamma(D_{s0}^{*+}\rightarrow (DK)^+)_{SU_f(4)}
\simeq 70$ MeV,  
%%%%%%%%%%%%%%%%%%%%%%%%%%%%%%%%%%%%%%%%%%%%%%%%%%%%%%%%%%%%%%%%%%%%%%%
where the asymptotic $SU_f(4)$ relation, Eq.~(\ref{eq:AME-SU(4)}), 
has been used. When we take account for the $SU_f(4)$ symmetry 
breaking as before, we obtain 
%%%%%%%%%%%%%%%%%%%%%%%%%%%%%%%%%%%%%%%%%%%%%%%%%%%%%%%%%%%%%%%%%%%%%%%
\begin{equation} 
 \Gamma_{D_{s0}^*}\sim 40 \,\,{\rm MeV}   \label{eq:width-D_{s0}}
\end{equation}
%%%%%%%%%%%%%%%%%%%%%%%%%%%%%%%%%%%%%%%%%%%%%%%%%%%%%%%%%%%%%%%%%%%%%%%
which is again rather broad in contrast with the four-quark $\hat F_I^+$ 
meson. Of course, the contribution of possible isospin non-conserving 
$D_{s0}^{*+}\rightarrow D_s^+\pi^0$ decays will be negligibly small, 
i.e., $\Gamma(D_{s0}^{*+}\rightarrow D_s^+\pi^0)_{SU_f(4)} \sim 0.6$ keV 
estimated in Eq.~(\ref{eq:rate-conv}). This is consistent with the fact 
that no scalar resonance has been observed in the region above the 
$D_{s0}^+(2317)$ resonance up to $\simeq 2.7$ GeV in the $D_s^+\pi^0$ 
mass distribution~\cite{BABAR-D_s}. It should be noted that the CLEO 
collaboration~\cite{CLEO-Kubota} have observed a peak around $2.39$ GeV 
in the $DK$ mass distribution but it has been taken away as a false peak 
arising from the decay, 
%%%%%%%%%%%%%%%%%%%%%%%%%%%%%%%%%%%%%%%%%%%%%%%%%%%%%%%%%%%%%%%%%%%%%
$D_{s1}(2536)\rightarrow D^*K\rightarrow D[\pi^0]K$,  
%%%%%%%%%%%%%%%%%%%%%%%%%%%%%%%%%%%%%%%%%%%%%%%%%%%%%%%%%%%%%%%%%%%%%
where the $\pi^0$ has been missed. However, we hope that it might  
involve true signals of a resonance corresponding to $D_{s0}^{*+}$ 
or that the resonance could be observed in the $DK$ channel in the 
region of $2.4 -2.5$ GeV by experiments with higher statistics and 
resolution. 

%%%%%%%%%%%%%%%%%%%%%%%%%%%%%%%%%%
\section{Summary}
%%%%%%%%%%%%%%%%%%%%%%%%%%%%%%%%%%

We have studied classification of charm scalar mesons and their decays 
into two pseudoscalar mesons and into the $D_s^*\gamma$. The two body 
decays has been calculated by using a hard pion (or kaon) technique 
in the infinite momentum frame which is an innovation of the old 
current algebra. The radiative decays have been studied under the 
vector meson dominance hypothesis. When the new resonance 
$D_{s0}^+(2317)$ is assigned to the $I_3=0$ component $\hat F_I^+$ of 
the iso-triplet four-quark mesons, the 
$\hat F_I^+\rightarrow D_s^+\pi^0$ decay can proceed through isospin 
conserving strong interactions. 
Therefore, the radiative $\hat F_I^+\rightarrow D_s^{*+}\gamma$ decay is 
much weaker than the $\hat F_I^+\rightarrow D_s^+\pi^0$. On the other 
hand, when it is assigned to an iso-singlet state (the four-quark 
$\hat F_0^+\sim [cn][\bar s\bar n]_{I=0}$ or the conventional scalar 
$D_{s0}^{*+}\sim \{c\bar s\}$), its decay into the $D_s^+\pi^0$ final 
state does not conserve isospin. Such an isospin non-conservation 
has been assumed to proceed through the $\pi^0$-$\eta$ mixing as usual. 
We have used the value of the isospin violating parameter 
$\epsilon \simeq 0.0105$ which was estimated long time ago and is of 
the order of the fine structure constant $\alpha$, i.e., 
$\epsilon \sim O(\alpha)$. Therefore the isospin non-conserving decays 
should be more strongly suppressed than the radiative decays. 
In this way, we have seen that the experimental constraint on the ratio 
of the decay rates for the $D_{s0}^+(2317) \rightarrow D_s^{*+}\gamma$ to 
the $D_{s0}^+(2317) \rightarrow D_s^+\pi^0$ favors 
the assignment of $D_{s0}^+(2317)$ to the iso-triplet four-quark 
meson $\hat F_I^+$. However, it is difficult to reconcile its assignment 
to an iso-singlet state with the constraint, even if the calculated mass 
values of the iso-singlet states ($D_{s0}^{*+}$ and $\hat F_0^+$) could  
reproduce well~\cite{MM,Narison,YSKK} the observed one, 
$m_{D_{s0}}=2317.4\pm 0.9$ MeV. Therefore, $D_{s0}^+(2317)$ can be 
considered as an evidence for existence of four-quark mesons with charm. 

Conventional charm scalar mesons also have been studied in relation to  
the broad enhancements in the $D\pi$ mass distributions which have been 
observed by the BELLE and by the FOCUS collaboration, independently. 
We have pointed out that each bump is unlikely to be saturated by a 
single scalar $D_0^*\sim \{c\bar n\}$ state, and we expect that each 
enhancement has a structure including at least two peaks, one arising 
from the four-quark $\hat D\sim [cn][\bar u\bar d]$ and the other from  
the conventional $D_0^*\sim\{c\bar n\}$, although the experiments have 
claimed that each bump is saturated by a conventional scalar meson. 
By comparing the decays of $D_0^*$ with the well-known 
$K_0^*(1430)\rightarrow K\pi$, the widths of $D_0^*$ have been predicted 
to be rather broad, $\Gamma_{D_0^*}\sim 90$ MeV (or $\sim 50$ MeV when 
the asymptotic $SU_f(4)$ symmetry breaking has been taken into account), 
but it is not enough to saturate the whole bump. In comparison, 
the four-quark $\hat D$ mesons have been expected  to have a width of 
at most $\sim 5 - 10$ MeV. Therefore we have proposed that each of the 
observed enhancements will consist of a broader $D_0^*$ in the region of 
its upper half and a narrower $\hat D$ in the region of its lower tail. 

The strange counterpart $D_{s0}^*\sim \{c\bar s\}$ of the conventional 
scalar $D_0^*$ is expected to have a mass around 
$m_{D_{s0}^*}\sim 2.45$ 
GeV from many different approaches. Its width is expected to be 
approximately saturated by the $D_{s0}^{*+}\rightarrow (DK)^+$ decays 
and is predicted to be $\sim 70$ MeV (or $\sim 40$ MeV when the 
asymptotic $SU_f(4)$ symmetry breaking has been taken into account). 
Therefore, we hope that experiments will observe a rather broad peak 
around $\sim 2.4 - 2.5$ GeV in the $DK$ mass distribution. 

One of remaining problems is that the neutral and doubly charged 
partners, $\hat F_I^0$ and $\hat F_I^{++}$, of $\hat F_I^+$ have not 
been observed. To solve this problem, we need to know production 
mechanism of $\hat F_I^{0,+,++}$ mesons. It will be one of 
our future projects. 

We emphasize that the values of the masses and widths of the scalar 
resonances which have been estimated in this article should not be taken 
too strictly since the width of $D_{s0}^+(2317)$ which has been used as 
the input data is narrow but its absolute value is still uncertain, and 
since possible mixing between $D_0^*$ and $\hat D$ through their common 
decay channels which might have considerable effects on the masses and 
widths of the mixed states~\cite{ABS} has been neglected. Such a mixing 
will depend on the details of hadron dynamics including four-quark 
mesons, and it will be a subject for future studies. 

Finally, we point out that it is desirable that the measured broad bumps 
in the $D\pi$ mass distributions be (re)analyzed by using an amplitude 
including at least two scalar resonances. It is also expected that the 
charm-strange scalar, $D_{s0}^{*+}\sim\,^3P_0\,\,\{c\bar s\}$, will be 
observed in the $DK$ channels by experiments with high statistics and 
resolution. 

\section*{Acknowledgments}
The author would like to thank Professor T.~Onogi and Dr. A.~Hayashigaki 
for valuable discussions. 
He also would like to appreciate Professor T.~Kunihiro and other members 
of nuclear theory group, Yukawa Institute for Theoretical Physics,  
and Professor H.~Suganuma, Physics Department of Kyoto University for 
discussions and comments. 
This work is supported in part by the Grant-in-Aid for Science Research, 
Ministry of Education, Science and  Culture, Japan (No. 13135101 and 
No. 16540243).

%%%%%%%%%%%%%%%%%%%%%%%%%%%%%%%%%
%%%%%%%%%%%%%%%%%%%%%%%%%%%%%%%%%

%\end{references}
%%%%%%%%%%%%%%%%%%%%%%%

\begin{thebibliography}{99}
\bibitem{afs} 
Roughly speaking, asymptotic flavor symmetry means flavor symmetry 
of asymptotic matrix elements (matrix elements taken between  
single hadron states) of charges, currents, etc. Asymptotic symmetry 
and its fruitful results are reviewed comprehensively in 
Ref.~\cite{suppl}.
\end{thebibliography}
\end{document}